\documentclass[12pt]{article}
\pdfoutput = 1
\usepackage{subcaption}
\usepackage{amsmath,amssymb,bm,epsfig,afterpage}
\usepackage{cite}
\usepackage{here}
\usepackage{color}
\usepackage{tabularx}
\usepackage{graphicx}
\usepackage{hyperref}
\usepackage{ulem}
\usepackage{authblk} 
\newcommand{\br}[2]{\text{Br}({#1}\to{#2})}

\newcommand{\eps}{\epsilon}

\newcommand{{\lag}}{\mathcal{L}}

\newcommand{\Lcal}{\mathcal{L}}
\newcommand{\Ncal}{\mathcal{N}}

\newcommand{\Pcal}{\mathcal{P}}
\newcommand{\Mcal}{\mathcal{M}}

\newcommand{\Ecal}{\mathcal{E}}

\newcommand{\la}{{\lambda}}

\newcommand{\order}[1]{\mathcal{O}\left({#1}\right)}
 
\newcommand{\pr}{\prime}
\newcommand{\Eps}{\mathcal{E}}
\newcommand{\define}{=}

\newcommand{\abs}[1]{\left|{#1}\right|}
\newcommand{\ol}[1]{\overline{#1}}

\newcommand{\MeV}{\mathrm{MeV}}
\newcommand{\GeV}{\mathrm{GeV}}
\newcommand{\TeV}{\mathrm{TeV}}
\newcommand{\rep}[1]{\mathbf{#1}}
\newcommand{\SM}{{\mathrm{SM}}}
\newcommand{\gmt}{g\mathrm{-}2}
\def\lsim{\raise0.3ex\hbox{$\;<$\kern-0.75em\raise-1.1ex\hbox{$\sim\;$}}}
\def\gsim{\raise0.3ex\hbox{$\;>$\kern-0.75em\raise-1.1ex\hbox{$\sim\;$}}}
\newcommand\Tstrut{\rule{0pt}{2.6ex}} % = `top' strut
\newcommand{\JK}[1]{\textcolor{red}{#1}}
\newcommand{\WA}[1]{\textcolor{blue}{#1}}
\allowdisplaybreaks[3]
 
\newcolumntype{Y}{&gt;{\centering\arraybackslash}X}

\numberwithin{equation}{section}

\begin{document}
%%%%%%%%%%%%%%%%
\begin{titlepage}
\begin{flushright}
 {\tt
CTPU-PTC-23-32 
}
\end{flushright}
\vspace{1.2cm}
\begin{center}
{\Large{\bf 
Semi-visible dark photon in a model with vector-like leptons 
for the $(g-2)_{e,\mu}$ and $W$-boson mass anomalies\\
\vspace{0.5cm}
}}
\vskip 2cm
%Waleed Abdallah, Mustafa Ashry, Junichiro Kawamura and Ahmad Moursy. 
Waleed Abdallah$^a$~\footnote{awaleed@sci.cu.edu.eg},
Mustafa Ashry$^a$~\footnote{mustafa@sci.cu.edu.eg}, 
Junichiro Kawamura$^{b}$~\footnote{junkmura13@gmail.com},
Ahmad Moursy$^c$~\footnote{a.moursy@fci-cu.edu.eg} 
\vskip 0.5cm
{\it $^a$ \normalsize Department of Mathematics, Faculty of Science, Cairo University, Giza 12613, Egypt}\\[3pt]
{\it $^b$ \normalsize Center for Theoretical Physics of the Universe, Institute for Basic Science (IBS), Daejeon 34051, Korea}\\[3pt]
{\it $^c$ \normalsize Department of Basic Sciences, Faculty of Computers and Artificial Intelligence, Cairo University, Giza 12613, Egypt}\\[3pt]
\vskip .5in

%%%%%%%%%%%%%%%%
\begin{abstract}
We propose a model realizes that a semi-visible dark photon 
which can contribute to the anomalous magnetic moment ($g-2$) 
of both electron and muon. In this model, 
the electron $g-2$ is deviated from the Standard Model (SM) prediction 
by the 1-loop diagrams involving the vector-like leptons, 
while that of muon is deviated due to a non-vanishing gauge kinetic mixing with photons.
We also argue that the $W$-boson mass can be deviated from the SM prediction 
due to the vector-like lepton loops, 
so that the value obtained by the CDF~II experiment can be explained. 
Thus, this model simultaneously explains the recent three anomalies in 
$g-2$ of electron and muon as well as the $W$-boson mass. 
The constraints on the $\mathcal{O}(1)~\mathrm{GeV}$ dark photon 
can be avoided because of the semi-visible decay of the dark photon, 
$A^\prime \to 2 N \to 2\nu \,2\chi \to 2\nu \,4e$, 
where $N$ is a SM singlet vector-like neutrino and 
$\chi$ is a CP-even Higgs boson of the $U(1)^\prime$ gauge symmetry.
\end{abstract}
\end{center}
%%%%%%%%%%%%%%%%
\end{titlepage}
%%%%%%%%%%%%%%%
\setcounter{footnote}{0}

\tableofcontents
\clearpage

%%%%%%%%%%%%%%%%%%%%%%
\section{Introduction}
%%%%%%%%%%%%%%%%%%%%%%

There is the long-standing discrepancy in the anomalous magnetic moment ($\gmt$) 
of muon between the Standard Model (SM) prediction~\cite{Aoyama:2020ynm,Aoyama:2012wk,Aoyama:2019ryr,Czarnecki:2002nt,Gnendiger:2013pva,Davier:2017zfy,Keshavarzi:2018mgv,Colangelo:2018mtw,Hoferichter:2019gzf,Davier:2019can,Keshavarzi:2019abf,Kurz:2014wya,Melnikov:2003xd,Masjuan:2017tvw,Colangelo:2017fiz,Hoferichter:2018kwz,Gerardin:2019vio,Bijnens:2019ghy,Colangelo:2019uex,Blum:2019ugy} 
and the experimental measurement~\cite{Bennett:2004pv,Abi:2021gix}. 
The latest world average of $\Delta a_\mu$ reports the $5.1\sigma$ discrepancy~\cite{Muong-2:2023cdq}, 
\begin{align}
\label{eq-delamuExp}
 \Delta a_\mu := a_\mu^{\mathrm{exp}} - a_\mu^{\SM} = 2.49~(48)\times 10^{-9}. 
\end{align} 
Whereas the recent lattice calculation~\cite{Borsanyi:2020mff} 
and the experiment determination~\cite{CMD-3:2023alj} 
of the hadron vacuum polarization contribution to the muon $\gmt$ 
point the value closer to the SM prediction, 
and hence the tension relaxes to a few sigma level. 
Nonetheless, we shall assume that the discrepancy is given by 
Eq.~\eqref{eq-delamuExp}, since the current situation is not conclusive. 
The electron $\gmt$ may also deviate from the SM prediction according
to the recent precise measurement of the fine structure constant 
using Cs atoms~\cite{Parker:2018vye}, 
and the discrepancy is given by~\cite{Davoudiasl:2018fbb}
\begin{align}
\label{eq-delaelExp}
 \Delta a_e := a_e^{\exp} - a_e^{\SM} = -8.7~(3.6)\times 10^{-13}, 
\end{align} 
and hence there is a $2.4\sigma$ discrepancy 
from the experimental value~\cite{Hanneke:2008tm,Hanneke:2010au}. 
Similarly to the muon $\gmt$, however, the situation is not conclusive 
because the fine structure constant determined by Rb atoms 
shows the value consistent with the SM~\cite{Morel:2020dww}. 
Nonetheless, we also assume that there is the discrepancy in Eq.~\eqref{eq-delaelExp}, 
especially the negative sign of its discrepancy. 
Simultaneous explanations for both anomalies have been studied 
in Refs.~\cite{Giudice:2012ms,Davoudiasl:2018fbb,Crivellin:2018qmi,Liu:2018xkx,
Dutta:2018fge,Parker:2018vye,Han:2018znu,Endo:2019bcj,
Bauer:2019gfk,Badziak:2019gaf,CarcamoHernandez:2019ydc,Cornella:2019uxs,Abdallah:2020biq,Ashry:2022maw,Ali:2021kxa,Cao:2023juc}.

The model with a $U(1)^\prime$ gauge symmetry and the vector-like fourth family 
is studied in Refs.~\cite{Kawamura:2019rth,Kawamura:2019hxp}~\footnote{
Other types of models with vector-like fermions and a $U(1)^\prime$ are studied 
in Refs.~\cite{Allanach:2015gkd,Altmannshofer:2016oaq,Megias:2017dzd,Raby:2017igl,Darme:2018hqg,Kulkarni:2023fyq,Lee:2021gnw,Lee:2022nqz,Dinh:2020inx,Zhou:2022cql,De:2021crr,Borah:2021khc}. 
}, 
to explain the muon $\gmt$ 
and another anomaly in the $b\to s\ell\ell$ process~\cite{Aebischer:2019mlg,Alguero:2019ptt,Alok:2019ufo,Ciuchini:2019usw,Datta:2019zca,Kowalska:2019ley,Arbey:2019duh,Kumar:2019nfv,Hiller:2019mou,deGiorgi:2022xhr}~\footnote{
The recent measurement of $R_{K^{(*)}}$ shows the consistent value 
with the SM prediction~\cite{LHCb:2022zom}. 
}.
In these works, the $U(1)^\prime$ gauge boson is assumed 
to be heavier than $100~\GeV$, so the gauge boson is called a $Z^\prime$-boson. 
The muon $\gmt$ is explained by the 1-loop diagrams involving the vector-like leptons 
via mixing with muons. 
In this case, however, the electron $\gmt$ can not be explained simultaneously 
because it causes the lepton flavor violations if the mixing with electrons is introduced. 
In Ref.~\cite{Kawamura:2022fhm}, 
it has been shown that the $W$-boson mass measured by the CDF~II~\cite{CDF:2022hxs},
\begin{align}
 m_W^{\mathrm{CDF}} = 80.4335~(94)~\GeV, 
\end{align}
which is larger than the previous measurements 
$m_W^{\mathrm{PDG}} = 80.379~(12)~\GeV$
and the SM prediction $m_W^\SM = 80.361~(6)~\GeV$~\cite{ParticleDataGroup:2020ssz}, 
can be explained by the 1-loop diagrams involving the vector-like leptons lighter 
than about $200$~GeV. 

In this work, we study a new parameter space of the model proposed 
in Ref.~\cite{Kawamura:2019rth,Kawamura:2019hxp}, 
where the $U(1)^\prime$ gauge boson is much lighter than the $Z$-boson mass and therefore we call it a dark photon $A'$ throughout this work. 
In such a scenario, the dark photon can explain $\Delta a_\mu$ 
if it is lighter than $\order{1}$~GeV 
and its gauge kinetic mixing with the photon is $\order{10^{-5}-10^{-2}}$ depending 
on the dark photon mass~\cite{Pospelov:2008zw}.
Note that the dark photon contribution from the gauge kinetic mixing 
can not explain the negative shift of the electron $g-2$ in Eq.~\eqref{eq-delaelExp}, 
since it is predicted to be positive. 
In this model, we can explain $\Delta a_e$ by the 1-loop diagrams 
involving the vector-like leptons as for $\Delta a_\mu$ in the heavy $Z^\prime$ scenario~\cite{Kawamura:2019rth,Kawamura:2019hxp}, without lepton flavor violations. 
We also point out that the $W$-boson mass measured by the CDF~II 
can be explained in the same manner as in Ref.~\cite{Kawamura:2022fhm}. 
Altogether, 
we study the light dark photon region of the model 
in Ref.~\cite{Kawamura:2019rth,Kawamura:2019hxp} 
in order to explain both electron and muon $\gmt$, 
as well as $m_W$ measured by the CDF~II experiment without extending the model. 

The dark photon explaining $\Delta a_\mu$ is excluded by the experiments 
if it decays dominantly to $e^+e^-$~\cite{BaBar:2014zli,NA482:2015wmo,LHCb:2017trq} 
or invisible particles~\cite{BaBar:2017tiz,NA64:2021xzo}. 
This limit will be relaxed and the dark photon explanation is still viable 
if the dark photon decays to both visible and invisible 
particles~\cite{Mohlabeng:2019vrz,Duerr:2019dmv,Duerr:2020muu,Abdullahi:2023tyk,NA64:2021acr}, 
namely if the dark photon is semi-visible. 
Interestingly, in this model, the SM singlet vector-like neutrino $N$ 
can be lighter than the dark photon,
and then $N$ can decay 
to the $U(1)^\prime$ breaking Higgs boson $\chi$
whose dominant decay mode is $e^+e^-$. 
Thus, the decay of the dark photon $A^\prime$ proceeds as 
$A^\prime \to 2 N \to 2\nu \,2\chi \to 2\nu \,4e$ which is a semi-visible decay.

The paper is organized as follows. In Sec.~\ref{Sec:model}, 
we briefly review the model with particular interests in the gauge kinetic mixing. 
We study the observables, 
including $\Delta a_{e}$, $\Delta a_\mu$ and $m_W$ in Sec.~\ref{Sec:gmin2}, 
and then discuss signals from the dark photon in Sec.~\ref{Sec:DPConst}. 
Finally, we draw our conclusions in Sec.~\ref{Sec:cnclsn}. 
The details of the model and the loop functions for the oblique parameters are 
respectively in Appendices~\ref{app-anal} and~\ref{app-loopfun}.

%%%%%%%%%%%%%%%%%%%%%%%%%%%%%%%%
\section{The model\label{Sec:model}}
%%%%%%%%%%%%%%%%%%%%%%%%%%%%%%%%

\begin{table}[t]
\centering
\begin{tabular}[t]{c|ccc|cccccc|c}
\hline\Tstrut %\hline\\\vspace{-.35in}\\
Gauge Symmetry & $\ell_{L_i}$ & $\ol{e}_{R_i}$ & $H$ & $L_L$ & $\ol{E}_R$ & $\ol{L}_R$ & $E_L$ & $\ol{N}_R$ & $N_L$ & $\Phi$ \\ \hline \hline 
$SU(2)_L$ & $\rep{2}$ & $\rep{1}$ & $\rep{2}$ & $\rep{2}$ & $\rep{1}$ & $\rep{2}$ & $\rep{1}$ & $\rep{1}$ & $\rep{1}$ & $\rep{1}$ \\
$U(1)_Y$ & $-1$ & $2$ & $-1$ & $-1$ & $2$ & $1$ & $-2$ & $0$ & $0$ & $0$ \\ \hline 
$U(1)^\prime$ & $0$ & $0$ & $0$ & $-1$ & $1$ & $1$ & $-1$ & $1$ & $-1$ & $-1$ \\
\hline
\end{tabular}
\caption{\label{Tab:MTCT}
Quantum numbers of the scalars and leptons in the model under 
the gauge symmetry $SU(2)_L\times U(1)_Y \times U(1)^\prime$. 
The index $i=1,2,3$ runs over the three generations of the SM leptons.
}
\end{table}

We review the model proposed in Refs.~\cite{Kawamura:2019rth,Kawamura:2019hxp} 
in which the SM is extended by a $U(1)^\prime$ gauge symmetry 
and a family of vector-like leptons. 
The matter contents of the model is summarized in Table~\ref{Tab:MTCT}.

\subsection{Gauge boson sector} 

Unlike the studies in Refs.~\cite{Kawamura:2019rth,Kawamura:2019hxp}, 
we explicitly introduce the gauge kinetic mixing 
of the $U(1)^\prime$ and $U(1)_Y$ symmetries. 
The gauge kinetic terms are given by 
\begin{align}
\label{Eq:gkmlg}
 \Lcal_{\mathrm{gauge}} 
= -\frac{1}{4} F_{\mu\nu} F^{\mu\nu} 
 -\frac{1}{4} F^\prime_{\mu\nu} F^{\prime\mu\nu} 
 -\frac{\eps}{2} F^\prime_{\mu\nu} F^{\mu\nu} 
 - \frac{1}{4} G^a_{\mu\nu} G^{a\mu\nu}, 
\end{align} 
where $F_{\mu\nu}$, $F^\prime_{\mu\nu}$ and $G^a_{\mu\nu}$ 
are the gauge field strengths of $U(1)_Y$, $U(1)^\prime$ and $SU(2)_L$, respectively. 
Here, $\eps$ is the gauge kinetic mixing factor. 
We denote the neutral vector fields of $U(1)_Y$, $U(1)^\prime$ and $SU(2)_L$
by $B_\mu$, $V_\mu$ and $W^3_\mu$, respectively. 
After the symmetry breaking by the SM Higgs boson 
and the $U(1)^\prime$ breaking scalar $\Phi$, 
the mass squared matrix for $(W^3_\mu, B_\mu, V_\mu)$ is given by 
\begin{align}
\label{eq-mVgauge}
 \Mcal_V^2 = m_W^2 
\begin{pmatrix}
 1 & -t_W & 0 \\ 
 -t_W & t_W^2 & 0 \\ 
 0 & 0 & t_V^2 \\ 
\end{pmatrix}, 
\end{align}
where $t_W := g_1/g_2$ and $t_V := m_V/m_W$ with 
$m_W := g_2 v_H/\sqrt{2}$ and $m_V = \sqrt{2} g^\prime v_\Phi$. 
Here, $g_1$, $g_2$ and $g^\prime$ are respectively the gauge coupling constants 
of $U(1)_Y$, $SU(2)_L$ and $U(1)^\prime$. 
The canonically normalized mass basis of the gauge bosons are defined as 
\begin{align}
 \begin{pmatrix}
 W^3_\mu \\ B_\mu \\ V_\mu 
 \end{pmatrix} 
=: 
\begin{pmatrix}
 s_W & c_W C_{WA^\pr} & c_W C_{WZ} \\ 
 c_W & -s_W C_{BA^\pr} & -s_W C_{BZ} \\ 
 0 & C_{VA^\pr} & C_{VZ}
\end{pmatrix} 
\begin{pmatrix}
 A_\mu \\ A^\prime_\mu \\ Z_\mu 
\end{pmatrix}, 
\end{align}
For $\eps, t_V \ll 1$, 
$C_{WA^\pr}$, $C_{BA^\pr}$, $C_{VZ} \sim \order{\eps}$ 
and 
$C_{WZ}$, $C_{BZ}$, $-C_{VA^\pr} \sim 1+\order{\eps^2}$. 
In this limit, 
\begin{align}
 m_{A^\prime}^2 \sim m_V^2(1+c_W^2\eps^2), \quad 
 m_Z^2 \sim \frac{m_W^2}{c_W^2} (1+s_W^2\eps^2), 
\end{align}
where $s_W := g_1/\sqrt{g_1^2+g_2^2} =: t_W c_W$. 
The explicit form of these matrices are shown in Appendix~\ref{app-anal}.

%%%%%%%%%%%%%%%%%%%%%%%%%%% 
\subsection{Fermion sector}
%%%%%%%%%%%%%%%%%%%%%%%%%%% 

In the gauge basis, 
the relevant part of the Lagrangian specifying the mass terms of the vector-like leptons and their Yukawa interactions 
are given by 
\begin{align}
\label{Eq:yukPhi}
\Lcal
\supset&\ 
m_L \ol{L}_R L_L + m_E \ol{E}_R E_L + m_N \ol{N}_R N_L \notag\\ 
&\ 
+ {\ol{e}_R}_i y^e_{ij} {\ell_L}_j H 
+ \Phi \la^L_{i}\ol{L}_R {\ell_L}_i - \Phi^* \la^E_{i} {\ol{e}_R}_i E_L \notag \\ 
&\ 
+\la_{e} \ol{E}_R L_L H - \la'_{e} \ol{L}_R \tilde{H} E_L 
+\la_{n} \ol{N}_R L_L \tilde{H} + \la'_{n} \ol{L}_R H N_L + h.c.
\end{align}
Here, $\tilde{H} \define i\sigma_2 H^*$ and $i,j=1,2,3$ label the SM generations.
After the symmetry breaking via non-zero vacuum expectation values (VEVs) of the scalar fields, 
$v_\Phi$ and $v_H$, the mass matrices
for 
$e_L = (e_{L_i}^-, L_L^-, E_L^-)$, $e_R = (e_{R_i}^-, E_R^-, L_R^-)$
and 
$n_L = (\nu_{L_i}, L_L^0, N_L)$, $n_R = (N_R, L_R^0)$
are given by
\begin{align}
 \Mcal_e = 
\begin{pmatrix}
 y_{ij} v_H	& 0	& \la_{L_i} v_\Phi \\ 
 0	& \lambda_{e}v_H	&	m_L \\ 
 \la_{E_j} v_\Phi	&	m_E	&	\lambda'_{e} v_H 
\end{pmatrix}, 
\quad 
 \Mcal_n = 
\begin{pmatrix}
 0	 & \la_{L_i} v_\Phi \\ 
 \lambda_{n}v_H & m _L \\ 
 m_N	& \lambda'_{n} v_H 
\end{pmatrix}. 
\end{align}
In this work, we do not explicitly introduce the right-handed neutrinos 
and treat neutrinos as massless particles. 
As shown in Ref.~\cite{Kawamura:2019hxp}, 
the phenomenology will not be changed up to $\order{v_H/M_{\mathrm{Maj}}}$, 
when we introduce the heavy right-handed neutrinos with Majorana mass 
$M_{\mathrm{Maj}} \sim 10^{10}~\GeV$. 
The mass matrices are diagonalized as 
\begin{align}
 U_{e_L}^\dag \Mcal_e U_{e_R} = 
\begin{pmatrix}
 m_{e_i} & 0 & 0 \\ 
 0 & m_{E_1} & 0 \\ 
 0 & 0 & m_{E_2} 
\end{pmatrix}, 
\quad 
 U_{n_L}^\dag \Mcal_n U_{n_R} = 
\begin{pmatrix}
 0 & 0 \\ 
 m_{N_1} & 0 \\
 0 & m_{N_2} \\ 
\end{pmatrix}, 
\end{align}
where $U_{e_{L,R}}$ and $U_{n_L}$ ($U_{n_R}$) are 
$5\times 5$ ($2\times 2$) unitary matrices. 
The leptons in the mass basis are defined as 
\begin{align}
 \hat{e}_A = U_{e_A}^\dag e_A, \quad 
 \hat{n}_A = U_{n_A}^\dag n_A, \quad 
A=L,R. 
\end{align}
The Dirac fermions are defined as
\begin{align}
\left[\psi_\ell\right]_J := 
\begin{pmatrix}
 [\hat{\ell}_L]_J \\ [\hat{\ell}_R]_J 
\end{pmatrix},
\quad 
\ell = e,n, 
\quad 
J = 1,2,3,4,5, 
\end{align}
where $[\hat{n}_R]_j = 0$ for $j=1,2,3$.

Throughout this work, we assume that the $U(1)^\prime$ breaking scalar $\Phi$ 
exclusively couples to the first generation, i.e. 
\begin{align}
\label{eq-lfvass}
 \la_{L_i} =: \la_L \delta_{1i} , 
\quad 
 \la_{E_i} =: \la_E \delta_{1i}, 
\quad 
 \la_{N_i} =: \la_N \delta_{1i}, 
\end{align}
so that the lepton flavor violations are not induced from the mixing. 
As we shall study the dark photon of $\order{1}~\GeV$, 
the VEV of $\Phi$ is expected to be in this order, 
which is much smaller than that studied in Refs.~\cite{Kawamura:2019rth,Kawamura:2019hxp}.
In this regime, with omitting the mixing with the second and third generations, 
the diagonalization matrices are approximately given by 
\begin{align}
 U_{e_L} =&\ 
\begin{pmatrix}
 1 & 0 & 0 \\ 
 0 & c_{e_L} & s_{e_L} \\ 
 0 &-s_{e_L} & c_{e_L} 
\end{pmatrix}
\begin{pmatrix}
 1-(\eta_{L_1}^2+\eta_{L_2}^2)/2 & \eta_{L_1} & -\eta_{L_2} \\ 
- \eta_{L_1} & 1 & 0 \\ 
 \eta_{L_2} & 0 & 1 \\ 
\end{pmatrix}, 
 \notag \\
 U_{e_R} =&\ 
\begin{pmatrix}
 1 & 0 & 0 \\ 
 0 & s_{e_R} & c_{e_L} \\ 
 0 & c_{e_R} & -s_{e_L} 
\end{pmatrix}
\begin{pmatrix}
 1-(\eta_{R_1}^2+\eta_{R_2}^2)/2 & -\eta_{R_1} & \eta_{R_2} \\ 
 \eta_{R_1} & 1 & 0 \\ 
 - \eta_{R_2} & 0 & 1 \\ 
\end{pmatrix}, 
\end{align}
where 
\begin{align}
 \eta_{L_1} := c_{e_R} \la_L \frac{v_\Phi}{m_{E_1}}, \quad 
 \eta_{L_2} := s_{e_R} \la_L \frac{v_\Phi}{m_{E_2}}, \quad 
 \eta_{R_1} := s_{e_L} \la_E \frac{v_\Phi}{m_{E_1}}, \quad 
 \eta_{R_2} := c_{e_L} \la_E \frac{v_\Phi}{m_{E_2}}. 
\end{align}
The first matrices diagonalize the right-lower $2\times 2$ block of $\Mcal_e$
and their analytical forms, as well as the diagonalization of the neutrino mass matrix, 
are shown in Appendix~\ref{app-anal}. 
The second matrices approximately diagonalize 
the small off-diagonal elements of the electron and the vector-like leptons 
up to the second order in $\eta := \order{\eta_{{L}_{1,2}},\eta_{R_{1,2}}}$.

%%%%%%%%%%%%%%%%%%%%%%%%%%%%%%%%%%%%%%%%%%%%%
\subsection{Fermion interactions}
%%%%%%%%%%%%%%%%%%%%%%%%%%%%%%%%%%%%%%%%%%%%%

The gauge interactions of the leptons with the neutral gauge bosons 
in the mass basis are given by 
\begin{align}
\Lcal_{VF} =&\ \sum_{\ell=e,n} \ol{\psi}_\ell \gamma_\mu \sum_{A=L,R} 
\Biggl[ 
e A^\mu Q_\ell \notag\\ 
& \quad 
+\sum_{X=A^\pr, Z} \frac{g_2}{c_W} X^\mu 
\left\{ I_{\ell_A} \left( c_W^2 C_{WX} + s_W^2 C_{BX} \right) 
 -s_W^2 Q_{\ell} C_{BX} + \frac{c_W g^\pr}{g_2} Q^\pr_{\ell_A} C_{VX} \right\} 
\Biggr] P_A \psi_\ell \notag\\ 
=: &\ 
- e A^\mu \ol{\psi}_e \gamma^\mu \psi_e 
+ \sum_{X=A^\pr, Z} \sum_{A=L,R} \sum_{\ell=e,n} 
X^\mu \ol{\psi}_\ell \gamma_\mu g^{X}_{\ell_A} P_A \psi_\ell,
\end{align}
where 
\begin{align}
I_{e_A} = -\frac{1}{2} U_{e_A}^\dag \Pcal_{A} U_{e_A} =: -\frac{1}{2}\Ecal_A, 
\quad 
I_{n_A} = +\frac{1}{2} U_{n_A}^\dag \Pcal_{A} U_{n_A} =: \frac{1}{2}\Ncal_A, 
\quad 
Q^\pr_{\ell_A} = - U_{\ell_A}^\dag \Pcal^\pr U_{\ell_A}, 
\end{align}
with $\Pcal_R := \mathrm{diag}(0,0,0,0,1) =: 1-\Pcal_L$ 
and $\Pcal^\pr := \mathrm{diag}(0,0,0,1,1)$. 
The electric coupling constant is defined as $e= g_1g_2/\sqrt{g_1^2+g_2^2}$, 
and the electric charged are $Q_e = -1$ and $Q_n = 0$. 
The $W$-boson couplings are given by
\begin{align}
\label{eq-Wdef}
 \Lcal_W = \frac{g_2}{\sqrt{2}} W^-_\mu \ol{\psi}_n \gamma^\mu \sum_{A=L,R} 
 h_A P_A \psi_e + h.c. 
 = \sum_{A=L,R} W^-_\mu \ol{\psi}_n \gamma^\mu g^W_{A} P_A \psi_e + h.c., 
\end{align}
where 
\begin{align}
h_A := U_{n_A}^\dag \Pcal_A U_{e_A}. 
\end{align}
The $U(1)^\prime$ Higgs boson $\Phi$ is expanded as 
\begin{align}
 \Phi = v_\Phi + \frac{1}{\sqrt{2}} \left(\chi + i a_\chi\right), 
\end{align}
where $a_\chi$ is the Nambu-Goldstone boson absorbed by the dark photon $A'$. 
The Yukawa interactions of the CP-even Higgs $\chi$ are given by 
\begin{align}
 -\Lcal_\chi = \frac{\chi}{\sqrt{2}} 
\sum_{\ell= e,n} \ol{\psi}_\ell Y^\chi_\ell P_L \psi_\ell + h.c., 
\end{align}
where 
\begin{align}
 Y_e^\chi = U^\dag_{e_L}
\begin{pmatrix}
 0 & 0 & \la_{L_i} \\ 
 0 & 0 & 0 \\ 
 \la_{E_j} & 0 & 0
\end{pmatrix}
U_{e_R}
, 
\quad 
 Y_n^\chi = U^\dag_{n_L}
\begin{pmatrix}
 0 & \la_{L_i} \\ 
 0 & 0 \\ 
 0 & 0
\end{pmatrix}
U_{n_R}.
\end{align} 
Up to $\order{\eps^2}$, the gauge couplings are given by 
\begin{align}
\label{eq-gZApApp}
g^Z_{\ell_A} \sim&\ 
\frac{g_2}{c_W}\left(I_{\ell_A} - s_W^2 Q_\ell\right) 
 + \eps s_W g^\pr Q^\pr_{\ell_A} 
 + \eps^2 g_2 t_W c_W \left\{ \frac{1}{2}I_{\ell_A} 
 - Q_\ell \left(1-\frac{s_W^2}{2}\right)\right\}, 
\\
g^{A^\prime}_{\ell_A} \sim&\ 
-g^\pr Q^\pr_{\ell_A} \left(1+\frac{c_W^2}{2}\eps^2\right) 
+ \eps c_W s_W g_2 Q_\ell. 
\end{align}
As explicitly shown in Appendix~\ref{app-anal}, we find 
\begin{align}
\label{eq-ELRapp}
 \Ecal_L \sim&\ 
\begin{pmatrix}
1-\eta_e^2/\la_E^2 & s_{e_L}\eta_e/\la_E & - c_{e_L} \eta_e/\la_E \\ 
 s_{e_L} \eta_e/\la_E & c_{e_L}^2 & c_{e_L}s_{e_L} \\ 
 -c_{e_L} \eta_e/\la_E & c_{e_L}s_{e_L}& s_{e_L}^2 
\end{pmatrix}, 
\quad 
 \Ecal_R \sim %&\ 
\begin{pmatrix}
\eta_e^2/\la_L^2 & c_{e_R} \eta_e/\la_L & -s_{e_R} \eta_e/\la_L \\ 
c_{e_R} \eta_e/\la_L & c_{e_R}^2 & -c_{e_R}s_{e_R} \\ 
-s_{e_R} \eta_e/\la_L & -s_{e_R}c_{e_R} & s_{e_R}^2 
\end{pmatrix}, 
\end{align}
where 
\begin{align}
 \eta_e:= \la_L \la_E v_\Phi
\left(\frac{s_{e_L}c_{e_R}}{m_{E_1}}+\frac{c_{e_L}s_{e_R}}{m_{E_2}}\right), 
\end{align}
will appear in $\Delta a_e$ expression in Sec.~\ref{Sec:gmin2}. 
For the $U(1)^\prime$ boson couplings, 
\begin{align}
\label{eq-QprApp}
 Q^\prime_{e_L}
\sim&\ 
\begin{pmatrix}
\eta_{L_1}^2+\eta_{L_2}^2 & -\eta_{L_1} & \eta_{L_2} \\ 
 -\eta_{L_1} & 1 & 0 \\
 \eta_{L_2} & 0 & 1 \\ 
\end{pmatrix}, 
\quad 
Q^\prime_{e_R}
\sim
\begin{pmatrix}
\eta_{R_1}^2+\eta_{R_2}^2 & \eta_{R_1} & -\eta_{R_2} \\ 
 \eta_{R_1} & 1 & 0 \\
 -\eta_{R_2} & 0 & 1 \\ 
\end{pmatrix}. 
\end{align}
Hence, the $Z$-boson couplings to the SM leptons 
are shifted at $\order{\eps^2, \eta^2}$
and those of the dark photon $A^\prime$ appears at $\eps$ 
with the sub-dominant contributions at $\order{\eps^2, \eta^2}$. 
The off-diagonal couplings of the SM leptons and the vector-like ones 
are induced at $\order{\eta}$. 
The structures are similar for the couplings involving the neutral leptons. 
The Yukawa couplings of the $\chi$ boson is approximately given by 
\begin{align}
 Y^\chi_e 
\sim 
\begin{pmatrix}
2 \eta_e & c_{e_R} \la_L & -s_{e_R} \la_L \\ 
-s_{e_L} \la_E & \order{v_\Phi/m_{E_1}} & \order{v_\Phi/m_{E_1}} \\
 c_{e_L} \la_E & \order{v_\Phi/m_{E_1}} & \order{v_\Phi/m_{E_2}} \\ 
\end{pmatrix}. 
\end{align}

%%%%%%%%%%%%%%%%%%%%%%%%%%%%%%%%%%%%%%%%%%%%%%%%%%%%
\section{Anomalous magnetic moments and $W$-boson mass} 
\label{Sec:gmin2}
%%%%%%%%%%%%%%%%%%%%%%%%%%%%%%%%%%%%%%%%%%%%%%%%%%%%

\subsection{Anomalous magnetic moments} 

\begin{figure}
\vspace{-.5in}
\centering
\includegraphics[trim={4.5cm 20.5cm 4cm 4cm},clip,scale=1.]{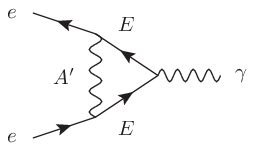}~\hspace{-2.9in}
\includegraphics[trim={4.5cm 20.5cm 4cm 4cm},clip,scale=1.]{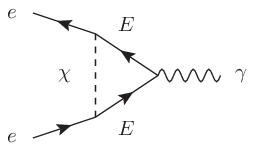}~\hspace{-2.9in}
\includegraphics[trim={4.5cm 20.5cm 4cm 4cm},clip,scale=1.]{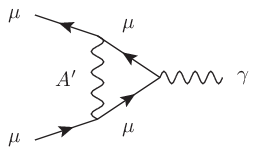}~\hspace{-2.9in}
\vspace{-.5in}
\caption{\label{Fig:gmin2feyndiag}
The Feynman diagrams dominantly contribute to the $\Delta a_e$ (left and middle) and $\Delta a_\mu$~(right). }
\end{figure}

The 1-loop contribution to the anomalous magnetic moment of the lepton $\ell = e,\mu$ 
via the neutral gauge boson $X=Z,A'$ and the charged leptons is given by 
\begin{align}
\label{eq:daMuZp}
 \delta_X a_{\ell} =& -\frac{m_\ell}{8\pi^2m_X^2} \sum_{B=1}^5 \left[
\left(\abs{\left[g^{X}_{e_L}\right]_{i_\ell B} }^2
 +\abs{\left[g^{X}_{e_R}\right]_{i_\ell B}}^2 \right)
 m_\ell F_Z(x^{X}_{e_B}) \right.\notag \\ 
&\hspace{5.0cm} \left.
+ \mathrm{Re}\left( \left[g^{X}_{e_L}\right]_{i_\ell B} 
 \left[{g^{X}_{e_R}}\right]^*_{i_\ell B} 
 \right) m_{e_B} G_Z(x^X_{e_B}) \right], 
\end{align}
where $x^{X}_{e_B}= m^2_{e_B}/m_{X}^2$. 
Here, $m_{e_B}$ is the mass of the $B$-th generation charged lepton, 
with flavor index $B=1,\dots,5$. The index $i_\ell = 1,2$ for $\ell = e, \mu$. 
The loop functions $F_Z(x), G_Z(x)$ are defined in Appendix~\ref{app-loopfun}.
The 1-loop contribution from the $\chi$ scalar to $\Delta a_\ell$ 
is given by~\cite{Dermisek:2013gta,Jegerlehner:2009ry}
\begin{align}
 \delta_\chi a_\ell = -
 \frac{m_\ell}{32\pi^2 m_\chi^2} \sum_{B=1}^5 & \left[
 \left(\abs{ \left[{Y^\chi_e}\right]_{i_\ell B}}^2
+\abs{\left[Y^\chi_e\right]_{Bi_\ell}}^2 \right) 
m_\ell F_S(y^\chi_{e_B}) \right. \notag \\
&\hspace{1.0cm} \left. + \mathrm{Re}\left({\left[Y^\chi_e\right]_{i_\ell B} 
 \left[Y^\chi_e\right]_{Bi_\ell}}\right)
 m_{e_B} G_S(y^\chi_{e_B})
\right],
\end{align}
where, $y^\chi_{e_B} := m_{e_B}^2/m_\chi^2$. Also, the loop functions $F_S(x), G_S(x)$ are defined in Appendix~\ref{app-loopfun}. 
Altogether, 
the new physics contribution to the anomalous magnetic moment is given by 
\begin{align}
 \Delta a_\ell = \delta_{A'} a_\ell + \delta_Z a_\ell + \delta_\chi a_\ell 
 - \delta_Z^\SM a_\ell , 
\end{align}
where the SM contribution via the $Z$-boson loop, 
\begin{align}
 \delta_Z^\SM a_\ell 
= -\frac{g_2^2 m_\ell^2}{8\pi^2 m_W^2} \left[ 
\left(\frac{1}{4}-s_W^2 + 2s_W^4\right) F_Z(x^Z_\ell) 
 + s_W^2\left(-\frac{1}{2}+s_W^2\right) G_Z(x^Z_\ell) 
 \right]
\end{align}
is subtracted. 
The contributions from the $Z$, $W$ and Higgs bosons are negligible 
because the off-diagonal couplings in the mass basis are suppressed. The Feynman diagrams dominantly contribute to $\Delta a_e$ and $\Delta a_\mu$ 
are shown in Fig.~\ref{Fig:gmin2feyndiag}.

\begin{figure}[t]
\centering
\begin{minipage}[t]{0.48\hsize}
 \includegraphics[width=0.95\hsize]{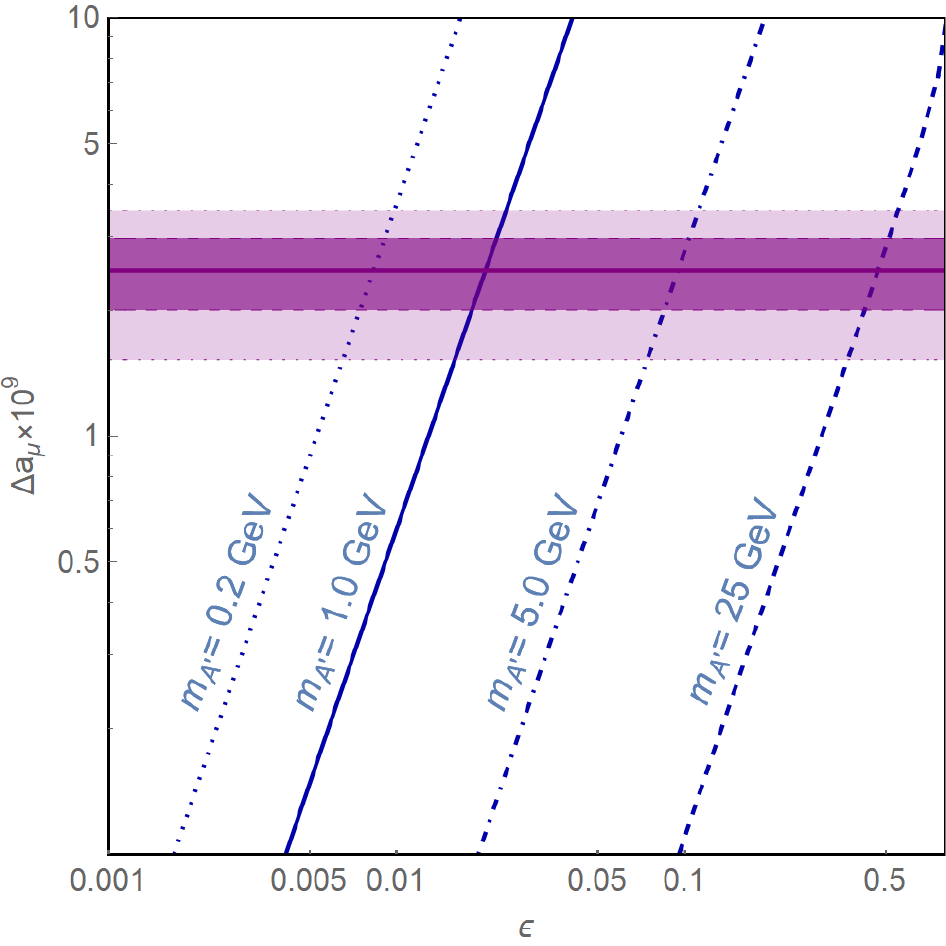}
\end{minipage}
\begin{minipage}[t]{0.48\hsize}
 \includegraphics[width=0.98\hsize]{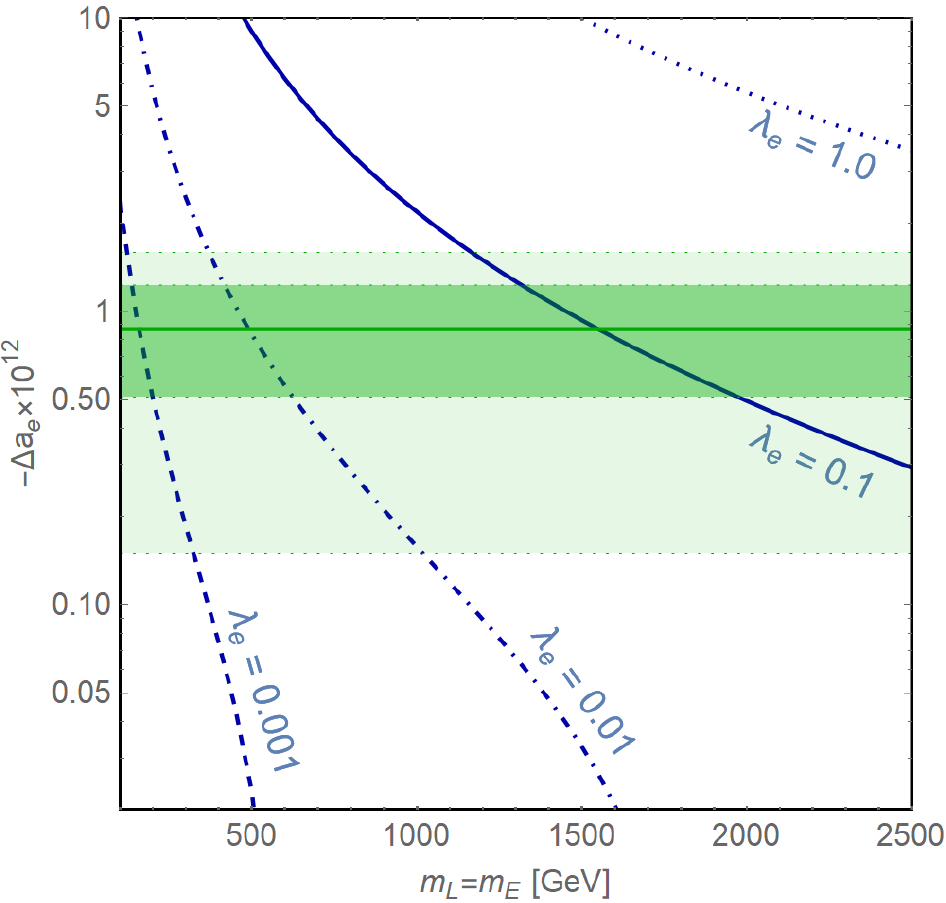}
\end{minipage}
\caption{
\label{fig-gm2s}
The left panel shows $\eps$ versus $\Delta a_\mu$ 
with $m_{A^\prime} = 0.2,1,5,25~\GeV$. 
The dark (light) purple region is the $1\sigma$~($2\sigma$) range. 
The right panel shows $m_L = m_E$ versus $-\Delta a_e$ 
with $\la_e = 0.001, 0.01, 0.1, 1.0$. 
The dark (light) green region is the $1\sigma$~($2\sigma$) range. 
The input parameters other than $m_L=m_E$ 
are chosen as the BP-A shown in Table~\ref{tab-BPs}. 
}
\end{figure}

Let us estimate the sizes of $\Delta a_\ell$ in our model. 
From Eq.~\eqref{eq-gZApApp}, 
the dark photon contribution to $\Delta a_\mu$ is approximately given by 
\begin{align}
\label{eq-dlaeps}
\Delta a_\mu \sim&\ \delta_{A^\prime} a_\mu \simeq 
-\frac{c_W^2s_W^2 g_2^2 m_\mu^2 \eps^2}{8\pi^2 m_{A^\prime}^2} 
 \left(2F_Z(x^{A^\prime}_\mu) + G_Z(x^{A^\prime}_\mu) \right) 
\\ 
\sim&\ 2.8\times 10^{-9} \times 
 \left(\frac{\eps}{0.02}\right)^2 \left(\frac{1~\GeV}{m_{A^\prime}}\right)^2 
 \left( \frac{2F_Z(x^{A^\prime}_\mu) + G_Z(x^{A^\prime}_\mu)}{-2/3} \right). 
\end{align}
Note that $2F_Z+G_Z$ is negative for $m_{A^\pr} > m_\mu$.

It is turned out that $\Delta a_e$ is dominantly 
from the 1-loop diagrams involving the vector-like leptons along with the dark photon or the $\chi$ boson, 
because of the chiral enhancement proportional to the vector-like lepton masses. 
From Eqs.~\eqref{eq-gZApApp} and~\eqref{eq-QprApp}, we find
\begin{align}
\label{eq-delaeApp}
 \Delta a_e \sim 
 -\frac{m_e \eta_e}{16\pi^2 v_\Phi} 
 \sim
-3.2\times 10^{-13} \times 
\left(\frac{1~\GeV}{v_\Phi}\right) \left(\frac{\eta_e}{10^{-7}}\right), 
\end{align}
and $\eta_e$ is approximately given by 
\begin{align}
\label{eq-etaApp}
 \eta_e \sim \la_L \la_E \frac{\la_e v_H v_\Phi}{m_L m_E}
 \sim 1.7 \times 10^{-7} \times \left(\frac{\la_e \la_L \la_E}{10^{-3}}\right)
\left(\frac{v_\Phi}{1~\GeV}\right) \left(\frac{10^3~\mathrm{GeV}}{\sqrt{m_Lm_E}}\right)^2, 
\end{align}
for $v_H \ll m_E$. 
Thus, the vector-like mass around the TeV-scale 
can explain the deviation in $\Delta a_e$ 
for the Yukawa coupling constants of $\order{0.1}$ 
and $v_\Phi \sim \order{1}~\GeV$. 
Note that the contribution from the gauge kinetic mixing will be sub-dominant 
when $\Delta a_\mu$ is explained 
because the coupling induced by the kinetic mixing is flavor universal 
and it is estimated as 
\begin{align}
\Bigl. \delta_{A^\prime} a_e \Bigr|_{\mathrm{\eps}}
= \frac{m_e^2}{m_\mu^2} \Delta a_\mu \simeq 5.8\times 10^{-14} \times 
 \left(\frac{\Delta a_\mu}{2.51\times 10^{-9}}\right). 
\end{align} 
For $\eta_e \sim 10^{-7}$, the $Z$-boson couplings of the SM leptons 
are very close to the SM one since the deviation is at $\order{\eta_e^2}$, 
see Eq.~\eqref{eq-ELRapp}.

Fig.~\ref{fig-gm2s} shows the values of $\Delta a_\mu$ (left) and $\Delta a_e$ (right) 
based on our numerical analysis. 
We see that $\Delta a_\mu$ is explained for $\eps \sim 0.02$ for the $1~\GeV$ dark photon 
as expected from Eq.~\eqref{eq-dlaeps}. 
For $(\eps, m_{A^\pr}) = (0.02, 1~\GeV)$, 
$\Delta a_e$ is explained by the vector-like lepton loops 
if the vector-like lepton masses are $1.5~\TeV$ ($500~\GeV$) with $\la_e = 0.1$~($0.01$), 
as expected from Eqs.~\eqref{eq-delaeApp} and~\eqref{eq-etaApp}. 
Thus, our model provides a unified explanation for both $\Delta a_e$ and $\Delta a_\mu$ 
without introducing lepton flavor violations.

\subsection{$W$-boson mass}
As shown in Refs.~\cite{Kawamura:2022uft,Kawamura:2022fhm},
the $W$-boson mass shift can be explained by the 1-loop effects 
of the fourth family vector-like leptons. The $T$ parameter~\cite{Peskin:1991sw,Peskin:1990zt} has a dominant contribution to this shift compared to the $S,U$ parameters and the $T$ parameter 
is given by~\cite{Lavoura:1992np,Dermisek:2022xal} 
\begin{align}
{16\pi s_W^2 c_W^2} T =&\ 
\sum_{a,\beta} \left\{ 
 \left( \abs{h^L_{a\beta}}^2 + \abs{h^R_{a\beta}}^2\right)\theta_+(y_a, y_\beta) 
 + 2\mathrm{Re}\left(h^L_{a\beta}h^{R*}_{a\beta}\right) \theta_-(y_a,y_\beta) 
 \right\}\notag \\ 
&\ - \sum_{a<b} \left\{ 
 \left( \abs{\Ncal^L_{ab}}^2 + \abs{\Ncal^R_{ab}}^2\right)\theta_+(y_a, y_b) 
 + 2\mathrm{Re}\left(\Ncal^L_{ab} \Ncal^{R*}_{ab}\right) \theta_-(y_a,y_b) 
 \right\} \notag \\
&\ 
-\sum_{\alpha<\beta} \left\{ 
 \left( \abs{\Ecal^L_{\alpha\beta}}^2 + \abs{\Ecal^R_{\alpha\beta}}^2\right)\theta_+(y_\alpha, y_\beta) 
 + 2\mathrm{Re}\left(\Ecal^L_{\alpha\beta}\Ecal^{R*}_{\alpha\beta}\right) 
 \theta_-(y_\alpha,y_\beta) 
 \right\}, 
\end{align} 
where the indices $a,b$ ($\alpha, \beta$) run over the neutral (charged) leptons, 
and $y_a := m_{e_a}^2/m_Z^2$, $y_\alpha := m_{n_\alpha}^2/m_Z^2$. 
Here, $h^A_{a\beta} = [h_A]_{a\beta}$, $\Eps^A_{\alpha\beta} = [\Eps_A]_{\alpha\beta}$
and $\Ncal^A_{ab} = [\Ncal_A]_{ab}$ for $A = L,R$. 
The formula of $2\pi S$ can be obtained by replacing 
$\theta_{\pm} \to \psi_{\pm}$ ($\theta_{\pm} \to \chi_\pm$) in the first line (the second and third lines), 
while by replacing $\theta_\pm \to \chi_{\pm}$ the formula of $-2\pi U$ can be obtained. 
The loop functions are defined in Appendix~\ref{app-loopfun}. % {app-loopfuns}. 
The $W$-boson mass is given by~\cite{Maksymyk:1993zm,Grimus:2008nb} 
\begin{align}
\hat{m}_W^2 
= &\ m_W^2\Bigr|_{\mathrm{SM}} \left[
1 + \frac{\alpha}{c_W^2-s_W^2} \left(
-\frac{S}{2} + c_W^2 T
 + \frac{c_W^2-s_W^2}{4s_W^2} U
\right) %
+ \Delta_W 
\right], 
\end{align}
where 
\begin{align}
 \Delta_W = \frac{c_W^2}{c_W^2-s_W^2} 
 \left(-\frac{\Delta m_Z^2}{m_Z^2} + t_W^2 \Delta h^L_{e\nu}\right), 
\end{align}
is the tree-level contribution from the $Z$-boson mass squared shift
$\Delta m_Z^2/m_Z^2 := m_Z^2/m_Z^2|_{\SM} - 1 \simeq s_W^2 \eps^2$
due to the kinetic mixing and the $W$-boson coupling to the SM leptons 
$\Delta h^L_{e\nu} := 1-[h_L]_{11}\sim\order{\eta^2}$~\footnote{
The tree-level contributions can be absorbed into the oblique parameters~\cite{Babu:1997st,Harigaya:2023uhg}, 
but our oblique parameters only include the loop effects from the vector-like leptons 
which are expected to be dominant. 
}. 
The tree-level contributions are too small to explain the shift in the $W$-boson mass, 
and hence $T\sim \order{0.1}$ is necessary to explain the CDF~II measurement. 
In fact, the limit on the dark-photon contributions to the EW precision data 
is $\eps < 2.7\times 10^{-2}$ for $m_{A^\pr} \ll 10~\GeV$~\cite{Curtin:2014cca}, 
where the most important effect is from the shift of the $Z$-boson mass 
which results the shift of the $W$-boson mass.

The $T$ parameter is approximately given by 
\begin{align}
 16\pi^2 c_W^2 s_W^2 T 
\simeq \frac{4(\la_n^\pr v_H)^4}{3m_{L}^2 m_Z^2} 
\left[1+\frac{1}{4}\left(\frac{\la_e^\pr m_{L}}{\la_n^\pr m_E}\right)^2 
 \left\{2-6\log\frac{m_E^2}{m_L^2}+3\left(\frac{\la^\pr_e}{\la^\pr_n}\right)^2\right\}
 \right], 
\end{align}
where we assume 
$m_N \ll v_H \ll m_L \ll m_E$ and $\la_e, \la_n \ll \la_e^\pr, \la_n^\pr$. 
The first term in the parenthesis 
comes from the $W$-boson contributions involving $N_2$ and $E_1$
which are sensitive to the mass difference in the doublet-like states. 
Since the second term is negative due to the logarithmic term, 
the $T$ parameter slightly increases as it is suppressed by $m_E$. 
For $m_L \ll m_E$, the $T$ parameter is estimated as 
\begin{align}
 T \sim 0.1\times\la_n^{\pr 4} 
 \left(\frac{230~\GeV}{m_L}\right)^2
% \left\{1-\frac{5}{4}\left(\frac{\la_e^\pr m_{L}}{\la_n^\pr m_E}\right)^2\right\}
. 
\end{align}
Thus, the shift of the $W$-boson mass suggested by the CDF~II measurement 
can be explained if $100 \lesssim m_L \lesssim 300~\GeV$ and $\la_n^\pr \sim 1$, 
so that the mass split between the neutral and charged doublet-like states 
is sizable~\footnote{
In models without the singlet vector-like neutrino $N$, 
the split should be originated from the charged leptons, 
and thus the charged vector-like lepton should be lighter than $200$ GeV 
to explain the CDF~II result~\cite{Kawamura:2022uft}.
}.

\begin{figure}[t]
\centering
\begin{minipage}[t]{0.48\hsize}
 \includegraphics[width=0.98\hsize]{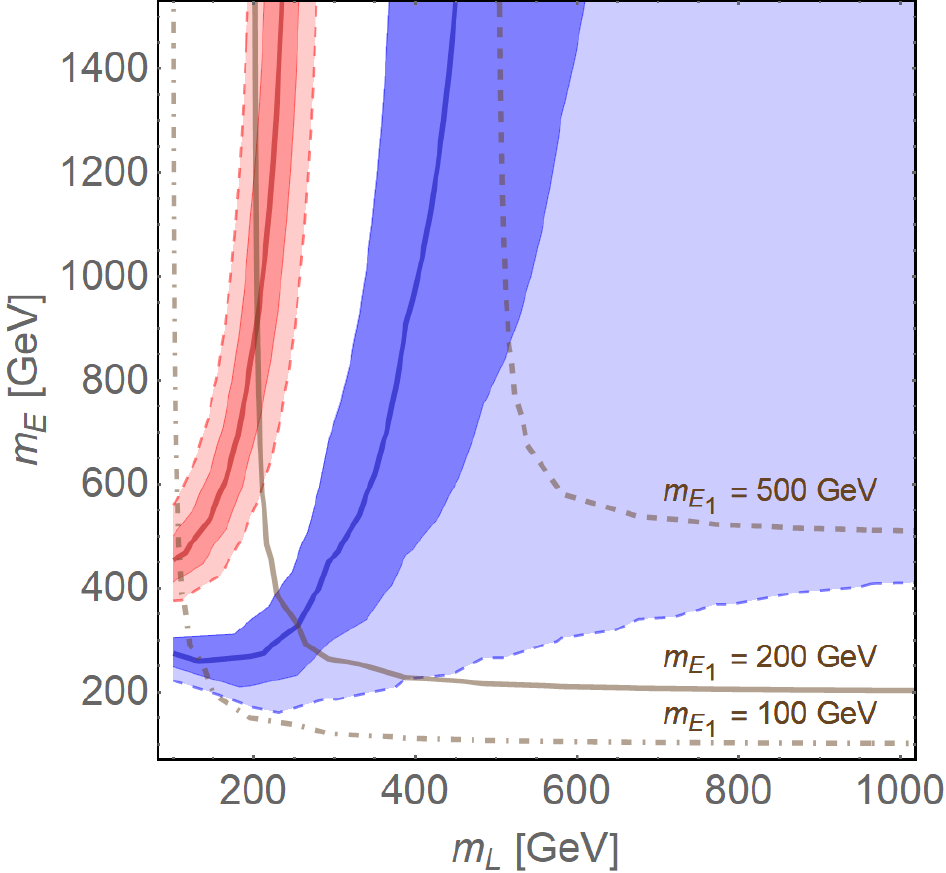}
\end{minipage}
\begin{minipage}[t]{0.48\hsize}
 \includegraphics[width=0.98\hsize]{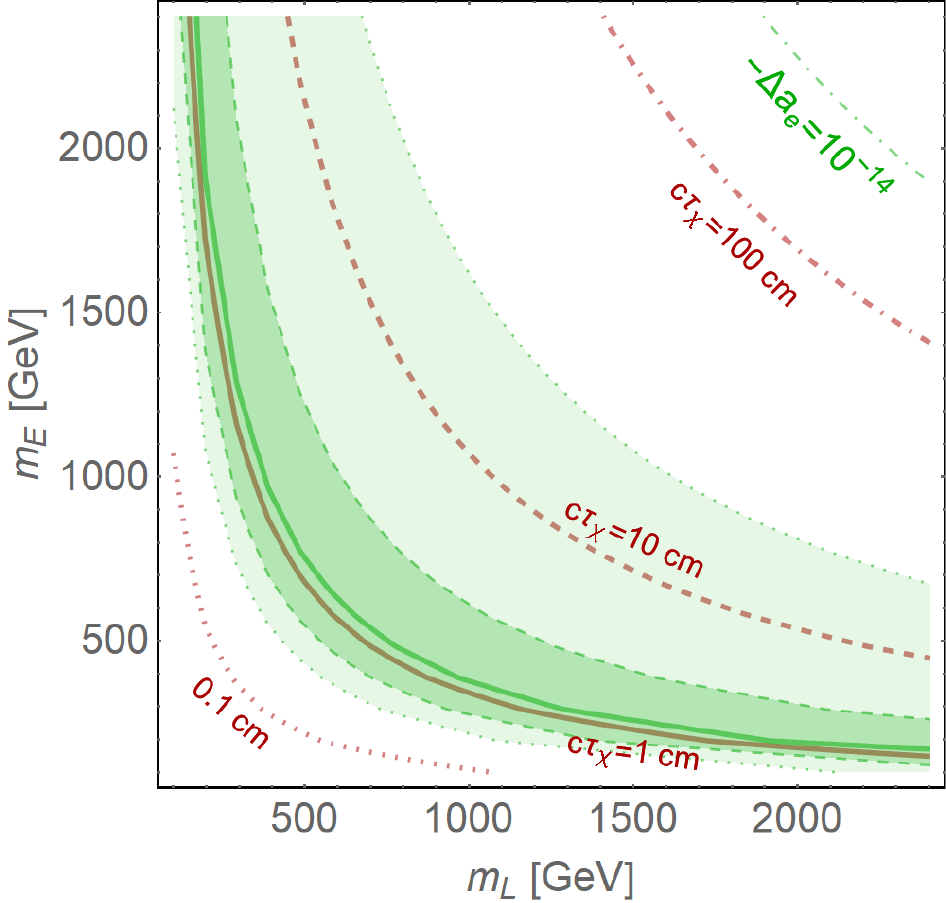} 
\end{minipage}
\caption{\label{fig-mWtauchi}
%(Left) $\Delta{a_\mu}$ and (right) $\Delta{a_e}$ versus $\epsilon$.
On the left panel, 
$m_W$ is explained the CDF~II and PDG result within the $1\sigma$ ($2\sigma$) range 
in the darker (lighter) red and blue regions, respectively. 
The solid line are the masses of the lightest charged exotic lepton 
$m_{E_1} = 100, 200$ and $500$ GeV from bottom to top (left to right). 
The value of $\la_L=\la_E$ chosen to explain $\Delta a_e$. 
On the right panel, $\Delta a_e = -8.7\times 10^{-13}$ on the green line, 
and it is within the $1\sigma$ ($2\sigma$) range 
in the darker (lighter) green region. 
The red lines are the length of flight of $\chi$. 
The inputs are those at the BP-B in Table~\ref{tab-BPs} 
except $(m_L, m_E)$ (and $\la_L=\la_E$) on the left (right) panel. 
}
\end{figure}

On the left panel of Fig.~\ref{fig-mWtauchi}, 
we plot the region where the $W$-boson mass is shifted 
due to the vector-like lepton loops.
The values favored by the CDF~II and PDG are explained in the $1\sigma$ ($2\sigma$) 
range in the darker (lighter) red and blue regions, respectively. 
In this plot, the input parameters except $m_L$, $m_E$ and $\la_L=\la_E$ 
are set to the values at the BP-B shown in Table~\ref{tab-BPs}. 
The value of $\la_L=\la_E$ are chosen to explain $\Delta a_e \simeq -8.7\times 10^{-13}$ 
based on the approximated formula in Eq.~\eqref{eq-delaeApp}, 
and hence both $\Delta a_e$ and $\Delta a_\mu$ are explained 
everywhere on the $(m_L, m_E)$ plane. 
The CDF~II value is explained 
if the doublet-like vector-like lepton is about $200~\GeV$, 
while that of the PDG is explained at $m_L \sim 500~\GeV$ depending 
on the singlet mass $m_E$. 
We shall briefly discuss about the LHC signals of the light vector-like charged leptons
in the next section.

\begin{table}[t]
\center
\caption{\label{tab-BPs}
Values of the inputs and the outputs at the benchmark points.
At the all points, the other inputs not shown in the table are set to
$\eps = 0.0203$, $\la_n = 0$
, $(m_{V}, m_\chi, m_{N_1}) = (1.0, 0.3, 0.4)~\GeV$
and $v_\Phi = 2\sqrt{2}~\GeV$.
The mass parameters are in the unit of GeV unless it is specified.
}
\begin{tabular}[t]{c|ccc} \hline
inputs & A & B & C \\ \hline\hline
$(m_L, m_E)$ & (1500., 1500.) & (300., 1400.) & (500., 1400.) \\
$\lambda_L=\lambda_E$ & 0.2 & 0.25 & 0.3 \\
$\lambda_e$ & 0.1 & 0.01 & 0.01 \\
$\lambda_e^\prime=\lambda_n^\prime$ & 0.5 & 1. & 1. \\ \hline
outputs & A & B & C \\ \hline\hline
$(m_{E_1}, m_{E_2})$ & (1448., 1553.) & (297.4, 1411.) & (495.4, 1412.) \\
$(m_{N_1}, m_{N_2})$ & (0.399, 1503.) & (0.346, 346.8) & (0.378, 529.4) \\ \hline
$-\Delta a_e\times 10^{13}$ & 9.326 & 7.698 & 6.557 \\
$\Delta a_\mu\times 10^{9}$ & 2.488 & 2.488 & 2.488 \\
$m_W$ & 80.3558 & 80.4046 & 80.3726 \\
$(S,T,U)$ & (2.388, 2.039, $-0.260$)$\times 10^{-4}$ & (0.012, 0.111, 0.009) & (0.007, 0.041, 0.002) \\\hline
$\Gamma_{A^\prime}$ [MeV] & 1.318 & 1.486 & 1.399 \\
Br$(A^\prime\to N_1N_1)$ & 0.9988 & 0.9989 & 0.9988 \\
$c\tau_{N_1}$ [cm] & 2.754 & 0.004444 & 0.006934 \\
Br$(N_1\to\chi\nu)$ & 1. & 1. & 1. \\
$c\tau_\chi$ [cm] & 1.078 & 1.541 & 2.065 \\\hline
Br$(E_1\to W N_1)$ & 0.7525 & 0.9216 & 0.9079 \\
Br$(E_1\to A^\prime e)$ & 0.1237 & 0.03918 & 0.04606 \\
Br$(E_1\to \chi e)$ & 0.1237 & 0.03918 & 0.04606 \\\hline
\end{tabular}
\end{table}

Table~\ref{tab-BPs} shows the three benchmark points (BPs)
which explain both $\Delta a_e$ and $\Delta a_\mu$. 
At the all points, $\eps=0.02$ and $m_{A^\prime} = 1~\GeV$ 
for $\Delta a_\mu \sim 2\times 10^{-9}$. 
The Yukawa couplings and vector-like masses are set to explain $\Delta a_e$. 
As discussed in the next section, 
we assume the spectrum $m_\chi < m_{N_1}/2 < m_{A^\prime}$ 
to realize the semi-visible dark photon compatible with the current limits. 
We also assume $\la_n \sim 0$ to keep $m_{N_1}$ of $\order{1}~\GeV$. 
At the BP-A, the vector-like leptons are about $1.5$ TeV, 
and hence the $W$-boson mass is very close to the SM value. 
At the BP-B (BP-C), the lightest charged lepton mass is about $300$ ($500$) GeV, 
so that the $W$-boson mass favored by the CDF~II (PDG) data is explained. 
We see that the $W$ mass shift is dominantly explained by the $T$ parameter,
and the other oblique parameters, $S$ and $U$, are much smaller.

%%%%%%%%%%%%%%%%%%%%%%%%%%%%%%%%%%
\section{Signals of light particles}
\label{Sec:DPConst}
%%%%%%%%%%%%%%%%%%%%%%%%%%%%%%%%%%

\subsection{Semi-visible dark photon}

\begin{figure}
%\vspace{-.5in}
\centering
\includegraphics[trim={4.5cm 20.5cm 4cm 4cm},clip,scale=1.]{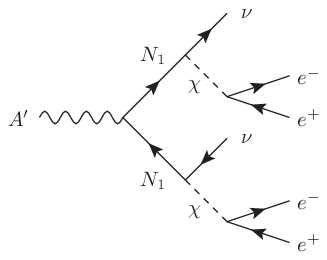}
%\vspace{-.5in}
\caption{\label{Fig:drkphtndcy}Dominant dark photon semi-visible decay.}
\end{figure}

The experiments exclude the dark photon responsible for the muon anomalous magnetic 
moment if it decays to a pair of electrons or invisible particles~\cite{BaBar:2014zli,NA482:2015wmo,LHCb:2017trq,BaBar:2017tiz,NA64:2021xzo}. 
The invisible dark photons are also searched in meson decays~\cite{BNL-E949:2009dza,NA62:2019meo,NA62:2020xlg,NA62:2020pwi}. 
There are limits from deep inelastic scatterings independently 
to decays of the dark photon, and the current limit for $\order{1}~\GeV$
dark photon is $\eps \lesssim 0.035$~\cite{Carrazza:2019sec,Kribs:2020vyk,Thomas:2021lub,McCullough:2022hzr,Thomas:2022qhj}, which is larger than our benchmark points $\eps = 0.02$. 
However, the experiments lose sensitivity for the other semi-visible dark photon 
decay modes, as discussed 
in Refs.~\cite{Mohlabeng:2019vrz,Duerr:2019dmv,Duerr:2020muu,Abdullahi:2023tyk}. 
There is the experimental analysis 
searching for such dark photon at the fixed-target experiment NA64~\cite{NA64:2021acr}. 
According to Refs.~\cite{NA64:2021acr,Abdullahi:2023tyk}, 
the dark photon explanation for $\Delta a_\mu$ is viable 
for $m_{A^\prime} \sim \order{0.1-1}~\GeV$ 
if the decay of heavy neutral fermion is fast enough. 
In our model, the dark photon will dominantly decay 
to a pair of vector-like neutrinos $N_1$ if $2m_{N} < m_{A^\prime}$. 
Then the vector-like neutrino $N_1$ will decay to the CP-even Higgs boson $\chi$ 
in the $U(1)^\prime$ breaking scalar $\Phi$. 
The scalar $\chi$ subsequently decays to a pair of electrons. 
Altogether, the decay chain of the dark photon is shown in Fig.~\ref{Fig:drkphtndcy}:
\begin{align}
\label{eq-AprDecay}
A^\prime\to N_1 N_1, 
\quad 
N_1 \to \nu \chi, 
\quad 
\chi \to e e, 
\end{align}
which is kinematically allowed if $m_{A^\prime}/2 > m_{N_1} > m_{\chi} > 2 m_e$. 
There are two pairs of electrons in the final state accompanied with two neutrinos. 
Thus, the signal at the experiments will be semi-visible 
if these decays happen inside detectors whose size is $\order{1\,m}$. 

The first decay $A^\prime \to N_1 N_1$ occurs promptly 
because $N_1 \sim N$ has the $U(1)^\prime$ charge and 
there is the coupling without suppression from $\eta$. 
The second decay $N_1 \to \chi \nu$ is relatively long, 
but is enough short since the coupling is suppressed only by $v_H/m_L$. 
Note that the decay width of $N_1$ is too small 
if the scalar $\chi$ is much heavier than $N_1$ 
so that there is only three-body decays via $A^\prime$ or the SM bosons. 
The decay width of the scalar $\chi$ is approximately given by 
\begin{align}
 \Gamma(\chi \to ee) = 
 \frac{m_\chi}{16\pi} \abs{\left[Y^\chi_e\right]_{ee}}^2 
 \left(1-\frac{4m_e^2}{m_\chi^2}\right)^{3/2}
\sim 
 \frac{m_\chi}{4\pi} \eta_e^2 
\end{align}
Interestingly, this is directly related 
to the approximated formula of $\Delta a_e$ in Eq.~\eqref{eq-delaeApp}, 
so that the length of flight of $\chi$ is estimated as 
\begin{align}
\label{eq-lifechi}
 c\tau_\chi \sim 1~\mathrm{cm} \times 
\left(\frac{8.8\times 10^{-13}}{\abs{\Delta a_e}}\right)^2 
\left( \frac{0.4~\GeV}{m_\chi}\right) 
\left( \frac{2\sqrt{2}~\GeV}{v_\Phi}\right)^2. 
\end{align} 
Thus, the scalar $\chi$ decays before reaching or inside the detectors 
if $\abs{\Delta a_e} \sim \order{10^{-13}}$, 
whereas the decay can not be detected 
and thus the signal is invisible if $\abs{\Delta a_e} \ll 10^{-13}$. 
The decay widths of $A^\prime$, $N_1$ and $\chi$ 
as well as the corresponding branching fractions at the BPs 
are shown in Table~\ref{tab-BPs}. 
We see that the lifetime of $A^\prime$ and $N_1$ are (much) less than $\order{\mathrm{cm}}$
and these dominantly decay to $N_1 N_1$ and $\chi \nu$, respectively. 
Here, we calculated the two-body decays of $A^\prime$ to two leptons 
and that of $N_1$ to $\nu \chi$ on top of the three body-decays 
via the gauge bosons which are negligibly small because of the suppressed couplings 
and the kinetic suppression. 
Thus, we confirmed that the dark photon decay 
can be dominated by $A^\pr \to N_1 N_1$, $N_1 \to \chi \nu$. 
If $\chi$ only decays to two electrons,
the length of flight is $\order{1}~\mathrm{cm}$, 
and hence this will be detected as prompt decay or displaced vertices 
depending on the detector design. 
It is also possible that the $\chi$ scalar 
decays to two pions if there are couplings in the quark sector as for the electrons.
In this case, the lifetime would be shorter. 
In Ref.~\cite{Abdullahi:2023tyk}, the dark photon decay proceeds as 
\begin{align}
 A^\prime \to \psi_i \psi_j, 
\psi_{i} \to \psi_{i-1} e^+ e^-, 
\psi_{j} \to \psi_{j-1} e^+ e^-, 
\cdots, 
\psi_{2} \to \psi_{1} e^+ e^-, 
\end{align}
where $\psi_{i}$'s are neutral exotic fermion 
and $\psi_1$ is considered to be stable, so that it can be the dark matter. 
In this scenario, the neutral fermion $\psi_i$ decays to three particles 
via off-shell dark photon, 
and thus their lifetimes tend to be longer than our case
in which the decay chain $N_1 \to \nu\chi, \chi \to ee$ proceeds via only two-body decays. 
Furthermore, the energy deposits from the $\chi$ decay will be larger 
than those from the decays of $\psi_i$ because of the larger phase space. 
Therefore, the signals form our dark photon will more easily 
evade from the experimental limits searching for invisible dark photons. 
We expect that the dark photon of $\order{0.1-1}~\GeV$ in our case
will not be excluded by the current data. 
The simulation as done in Ref.~\cite{Abdullahi:2023tyk} is beyond the scope of this work, but the simulation would confirm 
that the semi-visible dark photon responsible for the lepton magnetic moments 
would not be excluded by the experiments.

\subsection{The light vector-like neutrino and $U(1)'$ scalar}

In the realization of the semi-visible dark photon, 
the vector-like neutrino $N_1$ and the $U(1)^\pr$ scalar $\chi$ 
should also be $\order{0.1~\GeV}$. 
%In the literature, these light particles are dubbed 
%as the feebly interacting particles, 
%since their couplings to the SM particles should be feeble to be consistent 
%with the experimental data. 
The light vector-like neutrino $N_1$ mixes with the SM neutrinos 
through the mixing induced by $v_\Phi$ and $v_H$. 
Using the results in Appendix~\ref{app-anal}, 
the mixing between the light vector-like neutrino 
and the electron neutrino is approximately given by 
\begin{align}
\left[h_L\right]_{eN_1} \sim 
\frac{\la_L \la_e^{\pr 2} v_\Phi v_H^2}{2m_L m_E^2} 
\sim 
4\times 10^{-6} \times \la_e^{\pr 2}
\left(\frac{\la_L}{0.3}\right) 
\left(\frac{v_\Phi}{1~\GeV}\right) 
\left(\frac{500~\GeV}{m_L}\right) 
\left(\frac{1500~\GeV}{m_E}\right)^2, 
\end{align} 
where $h_L$ is defined in Eq.~\eqref{eq-Wdef}, 
and thus this mixing is $\order{10^{-6}}$ for our model. 
This is safely below the current experimental limits 
on the active-sterile mixing for $m_{N_1}\sim\order{0.1~\GeV}$, 
see Fig.~6 in Ref.~\cite{Bolton:2019pcu}.

In our model, the light scalar $\chi$ of mass $\sim\order{0.1~\GeV}$ is coupled to $e^+ e^-$ with a coupling strength is estimated to be $2\eta_e \sim \order{10^{-7}}$ from Eq.~\eqref{eq-etaApp}. 
Such a light scalar is constrained by the collider experimental limits searching 
for $e^+e^- \to \gamma \chi (\to e^+ e^-)$ 
at BaBar~\cite{BaBar:2014zli,Batell:2016ove}, KLOE~\cite{Anastasi:2015qla}, 
Belle-II projection~\cite{Batell:2017kty,Belle-II:2010dht,Belle-II:2018jsg} 
and the electron beam dump experiments~\cite{Liu:2016qwd,Batell:2017kty}. 
Relevant to the light scalar mass range under consideration, 
these experiments impose an upper bound on its coupling with an $e^+ e^-$ pair, 
$Y^\chi_e\lesssim 10^{-3}$. 
The limit of $Y^\chi_e \lesssim 10^{-3}$ is obtained for $m_\chi \gtrsim 20~\MeV$ 
from BaBar~\cite{BaBar:2014zli} 
and Belle-II~\cite{Batell:2017kty,Belle-II:2010dht,Belle-II:2018jsg}. 
The beam dump experiments~\cite{Riordan:1987aw,Bjorken:1988as,Davier:1989wz}
have sensitivities for $m_\chi \sim 1\mathrm{-}200~\MeV$ 
with $Y^\chi_e \sim 10^{-2}\mathrm{-}10^{-6}$, 
and no limits for heavier masses~\cite{Liu:2016qwd}. 
Therefore, our values are comfortably below this upper bound.

\subsection{Vector-like lepton search at the LHC}
\label{sec-VLLsearch}

We briefly discuss the LHC limits for the charged vector-like lepton 
$E_1$, which is expected to be light particularly to explain the $W$-boson mass shift. 
The vector-like leptons might be excluded by the LHC limits. 
For the doublet-like leptons, the mass below $800~\GeV$ is excluded if it decays to 
the SM particles~\cite{CMS:2022nty,CMS:2019lwf}. In our model, however, the vector-like lepton $E_1$ decays 
to $W N_1$, $A^\prime e$ and/or $\chi e$, 
as discussed in Refs.~\cite{Kawamura:2022fhm}. 
The branching fractions of these decay modes of our BPs are shown in Table~\ref{tab-BPs}. For the BPs, the dominant decay mode $E_1 \to W N_1$,
followed by $N_1 \to \chi \nu \to ee\nu$, 
has at least two electrons in the final states. 
This case might be covered by the same search studied in Ref.~\cite{Kawamura:2022fhm}, 
but there is no study for searching for the cascade decay. 
Thus, we can not exclude this possibility. In addition, due to the many-body decay cascade, the phase space of the decay $E_1 \to W N_1$ is small and thus the many leptons in the final state are relatively soft.
The sub-dominant decay modes $E_1 \to \chi e\to eee$ and $E_1 \to A' e\to eee$ have three electrons in the final state. 
These signals are similar to those from $E_1 \to Z^\prime \mu \to \mu\mu\mu$,
studied in Ref.~\cite{Kawamura:2022fhm},
which excludes the vector-like lepton masses up to 500 GeV 
for $\br{E_1}{ee e} \sim 10\%$. For our BP-A, $\br{E_1}{ee e} \simeq 12\%$ and $m_{E_1}\simeq 1.5$~TeV, which is safely above this limit.
On the other hand, the limit for branching fractions less than 10\% are not visible, therefore the BP-B and BP-C whose $\br{E_1}{ee e} \simeq 5\%$, may be allowed. We also note that this will not be the case 
if $\chi$ dominantly decays to quarks\footnote{If $\chi$ couples with quarks, 
the precision measurements of kaon decays will constrain the $\chi$ 
as discussed in Ref.~\cite{Liu:2020qgx}, depending on the flavor structure of the quark couplings. Also, the relation of the lifetime to $\Delta a_e$, in Eq.~\eqref{eq-lifechi} is changed by the mixing with quarks. 
A concrete study is beyond the scope of this paper.}.

%%%%%%%%%%%%%%%%%%%%%%%%%%%%%%%%%%%%%%%%%%%%
\section{Conclusions}
%%%%%%%%%%%%%%%%%%%%%%%%%%%%%%%%%%%%%%%%%%%%
\label{Sec:cnclsn}

In this work, 
we proposed a scenario in which 
both anomalies in electron and muon anomalous magnetic moments 
are explained without extending the model proposed 
in Refs.~\cite{Kawamura:2019rth,Kawamura:2019hxp}. 
The discrepancy for electron, $\Delta a_e$, is explained 
by the 1-loop diagrams involving the dark photon and the vector-like leptons, 
whereas that for muon, $\Delta a_\mu$ is explained 
by the 1-loop diagrams induced by the gauge kinetic mixing with photons. 
Since the latter effect is always positive, 
we can not consider the opposite case 
in which $\Delta a_e < 0$ is explained by the gauge kinetic mixing. 
Since two discrepancies are explained by the different origins, 
there is no lepton flavor violations induced by the new particles in the model. 
We also showed that the $W$-boson mass measured at the CDF~II can be explained 
if the vector-like lepton is below 300 GeV. 
Such a light vector-like lepton would be excluded 
by the high-multiplicity lepton channels at the LHC, 
depending on its decay modes, as discussed in Sec.~\ref{sec-VLLsearch}. 
If the light vector-like lepton is not excluded by the LHC, 
this model can address the three anomalies simultaneously.

The dark photon explanation of $\Delta a_\mu$ 
is severely constrained by the experiments in the simplest setups. 
In our model, however, the dark photon can decay to a pair of vector-like neutrinos, 
$A^\pr \to N_1N_1$, followed by the decays $N_1 \to \chi (\to ee)\nu$, 
so that the dark photon becomes semi-visible which is not excluded 
by the dark photon searches. 
We also find that the lifetime of the $\chi$ field is directly related 
to the new physics contribution to $\Delta a_e$, 
and thus our resolution to avoid the invisible dark photons search 
works only if $|\Delta a_e| \gtrsim 10^{-14}$. 
This scenario would be probed by the direct searches 
for the semi-visible dark photons, 
or pair productions of the charged vector-like leptons at the LHC, 
which are subjects of our future works. 
Our model provides an explicit example of the semi-visible dark photon 
relying only on two-body decays which are qualitatively different 
from those considered in the literature.

%%%%%%%%%%%%%%%%%%%%%%%%%%%
\section*{Acknowledgments} %CKNOWLEDGMENTS}
%%%%%%%%%%%%%%%%%%%%%%%%%%%
The work of J.K. is supported in part by
the Institute for Basic Science (IBS-R018-D1).

%%%%%%%%%%%
%\clearpage
\appendix
\section{Details of the model}
\label{app-anal}

\subsection{Diagonalization of vector boson mass matrix} 

We show the explicit form of the diagonalization matrix 
to obtain the canonically normalized mass basis of the vector bosons. 
We decompose the diagonalization matrix as 
\begin{align}
 \begin{pmatrix}
 W^3_\mu \\ B_\mu \\ V_\mu 
 \end{pmatrix} 
=: \mathcal{E} R_1 R_2 
\begin{pmatrix}
A_\mu \\ A^\prime_\mu \\ Z_\mu 
\end{pmatrix}, 
\end{align}
where $\mathcal{E}$ canonically normalizes the kinetic terms, 
$R_1$ block diagonalize the massless photon and the others, 
and $R_2$ diagonalize the $2\times 2$ block of the massive bosons. 
Their explicit forms are given by 
\begin{align}
\mathcal{E} = \frac{1}{\sqrt{2}} 
\begin{pmatrix}
 \sqrt{2} & 0 & 0 \\ 
 0 & \eta_+ & -\eta_- \\ 
 0 & \eta_+& \eta_- 
\end{pmatrix}, 
\ \
R_1 = \frac{1}{\sqrt{2}}
\begin{pmatrix}
 \sqrt{2} s_W & \sqrt{2} c_W & 0 \\ 
 c_W/\eta_+ & -s_W/\eta_+ & 1/\eta_- \\ 
 -c_W /\eta_- & s_W/\eta_- & 1/\eta_+ 
\end{pmatrix}, 
\ \ 
R_2 = 
\begin{pmatrix}
 1 & 0 & 0 \\ 
 0 & c_V & s_V \\ 
 0 & -s_V & c_V \\ 
\end{pmatrix}, 
\end{align}
where $\eta_\pm := 1/\sqrt{1\pm\eps}$ and 
\begin{align}
c_V := \sqrt{\frac{1}{2}\left(1-\frac{1-(1+s_W^2)\eps^2-c_W^2 t_V^2}{\sqrt{d_V}} \right) },
\quad 
s_V := \mathrm{sign}(\eps)
 \sqrt{\frac{1}{2}\left(1+\frac{1-(1+s_W^2)\eps^2-c_W^2 t_V^2}{\sqrt{d_V}} \right) },
\end{align}
with 
\begin{align}
 d_V := (1-\eps^2 c_W^2)^2
 - 2\left\{1-(1+s_W^2)\eps^2 \right\} c_W^2 t_V^2 
 +c_W^4t_V^4 . 
\end{align}
Altogether, the diagonalization matrix has the form 
\begin{align}
 \Eps R_1 R_2 = 
\begin{pmatrix}
 s_W & c_W c_V & c_W s_V \\ 
c_W & -s_W c_V + \eps s_V \eta_+\eta_- & 
-s_V s_W -\eps c_V \eta_+\eta_- \\ 
0 & -s_V \eta_+\eta_- & c_V \eta_+\eta_- 
\end{pmatrix}
=: 
\begin{pmatrix}
 s_W & c_W C_{WA^\pr} & c_W C_{WZ} \\ 
 c_W & -s_W C_{BA^\pr} & -s_W C_{BZ} \\ 
 0 & C_{VA^\pr} & C_{VZ}
\end{pmatrix}. 
\end{align}
The masses after diagonalization are given by 
 \begin{align}
 m_{A^\pr}^2 &= \frac{m_W^2}{2c_W^2(1-\eps^2)} 
 \left( 1 + c_W^2 (t_V^2 - \eps^2) - \sqrt{d_V} 
 \right), \\
 m_{Z}^2 &= \frac{m_W^2}{2c_W^2(1-\eps^2)} \left( 1 + c_W^2 (t_V^2 - \eps^2) 
 + \sqrt{d_V} 
 \right). 
\end{align}
Up to the second order in $\eps$ and $t_V$, 
\begin{align}
 C_{WA^\pr}\sim&\ s_W \eps, \quad 
 C_{BA^\pr}\sim -\frac{c_W^2}{s_W}\eps, \quad
 C_{VA^\pr}\sim -\left(1+\frac{c_W^2}{2}\eps^2\right) 
\\ \notag 
 C_{WZ}\sim&\ 1-\frac{s_W^2}{2} \eps^2, \quad 
 C_{BZ}\sim 1+\left(1-\frac{s_W^2}{2}\right)\eps^2, \quad 
 C_{VZ}\sim s_W\eps, 
\end{align}
and 
\begin{align}
 m_{A^\pr}^2 \sim m_V^2 (1+c_W^2 \eps^2), 
\quad 
 m_Z^2 \sim \frac{m_W^2}{c_W^2} \left(1+s_W^2 \eps^2\right). 
\end{align}

%%%%%%%%%%%%%%%%%%%%%%%%%%%%%%%%%%%%%%%%%%%%%%%%%%%%%%%%%%%%%
\subsection{Diagonalization of the fermion mass matrices} 
%%%%%%%%%%%%%%%%%%%%%%%%%%%%%%%%%%%%%%%%%%%%%%%%%%%%%%%%%%%%%

We show the diagonalization matrices of the leptons, 
\begin{align}
 \Mcal_e = 
\begin{pmatrix}
 y_{1} v_H	& 0	& \la_{L} v_\Phi \\ 
 0	& \lambda_{e}v_H	&	m_L \\ 
 \la_{E} v_\Phi	&	m_E	&	\lambda'_{e} v_H 
\end{pmatrix}, 
\quad 
 \Mcal_n = 
\begin{pmatrix}
 0	 & \la_{L} v_\Phi \\ 
 \lambda_{n}v_H & m _L \\ 
 m_N	& \lambda'_{n} v_H 
\end{pmatrix}, 
\end{align}
for $v_\Phi, m_N \ll v_H \lesssim m_L, m_E$. 
Here, we omit the second and third generations 
under the assumption of Eq.~\eqref{eq-lfvass}. 
We also assume that $m_L, m_E, m_N > 0$. 
The diagonalization matrices of the charged leptons are given by 
\begin{align}
 U_{e_L} =&\ 
\begin{pmatrix}
 1 & 0 & 0 \\ 
 0 & c_{e_L} & s_{e_L} \\ 
 0 &-s_{e_L} & c_{e_L} 
\end{pmatrix}
\begin{pmatrix}
 1-(\eta_{L_1}^2+\eta_{L_2}^2)/2 & \eta_{L_1} & -\eta_{L_2} \\ 
- \eta_{L_1} & 1 & 0 \\ 
 \eta_{L_2} & 0 & 1 \\ 
\end{pmatrix}, 
\\ \notag 
 U_{e_R} =&\ 
\begin{pmatrix}
 1 & 0 & 0 \\ 
 0 & s_{e_R} & c_{e_L} \\ 
 0 & c_{e_R} & -s_{e_L} 
\end{pmatrix}
\begin{pmatrix}
 1-(\eta_{R_1}^2+\eta_{R_2}^2)/2 & -\eta_{R_1} & \eta_{R_2} \\ 
 \eta_{R_1} & 1 & 0 \\ 
 - \eta_{R_2} & 0 & 1 \\ 
\end{pmatrix}, 
\end{align}
up to the second order in $\eta := \order{\eta_{{L}_{1,2}},\eta_{R_{1,2}}}$. 
The first matrices diagonalize the right-lower $2\times 2$ block of $\Mcal_e$. 
The angles are given by 
\begin{align}
\label{eq-cseLR}
 c_{e_L} =&\ \sqrt{\frac{1}{2}\left(1-\frac{T_{e_L}}{\sqrt{D_e}}\right)}, 
\quad 
 s_{e_L} = \sigma_{e_L} 
 \sqrt{\frac{1}{2}\left(1+\frac{T_{e_L}}{\sqrt{D_e}}\right)}, \\ \notag 
 c_{e_R} =&\ \sqrt{\frac{1}{2}\left(1+\frac{T_{e_R}}{\sqrt{D_e}}\right)}, 
\quad 
 s_{e_R} = -\sigma_{e_R}\sqrt{\frac{1}{2}\left(1-\frac{T_{e_R}}{\sqrt{D_e}}\right)}, 
\end{align}
where 
\begin{align}
 S_e :=&\ m_E^2 + m_L^2 + \left(\la_e^2 + \la_e^{\pr 2}\right)v_H^2, 
\quad 
 D_e := S_e^2 - 4 \left(m_L m_E - \la_e\la_e^\pr v_H^2\right)^2, 
\\ \notag 
 T_{e_L} :=&\ m_L^2 - m_E^2 + (\la_e^2-\la_e^{\pr 2})v_H^2, 
\quad 
 T_{e_R} := m_E^2 - m_L^2 + (\la_e^2-\la_e^{\pr 2})v_H^2, 
\end{align}
and 
\begin{align}
 \sigma_{e_L} := \mathrm{sign}\left( \la_e m_E + \la_e^\pr m_L \right), 
\quad 
 \sigma_{e_R} := \mathrm{sign}\left( \la_e m_L + \la_e^\pr m_E\right). 
\end{align}
The second matrices diagonalize the mixing between the first generation 
and the vector-like lepton. 
The singular values are given by~\footnote{
The diagonal elements after the rotation by the first matrix are given by, 
\begin{align}
 \mu_{E_1} =&\ \frac{\la_e v_H}{2c_{e_L}s_{e_R}} 
 \left[1-\frac{1}{\sqrt{D_e}} 
 \left\{S_e+2 \frac{\la_e^\pr}{\la_e} \left( m_Lm_E-\la_e\la_e^\pr v_H^2 \right) \right\}
 \right], 
\\ \notag 
 \mu_{E_2} =&\ \frac{\la_e v_H}{2s_{e_L}c_{e_R}} \left[1+\frac{1}{\sqrt{D_e}} 
 \left\{S_e+2\frac{\la_e^\pr}{\la_e} \left( m_Lm_E- \la_e\la_e^\pr v_H^2 \right)\right\}
\right], 
\end{align}
such that 
\begin{align}
 \begin{pmatrix}
 c_{e_L} & s_{e_L} \\ -s_{e_L}& c_{e_L}
 \end{pmatrix}
\begin{pmatrix}
 \la_e v_H & m_L \\ m_E & \la^\pr_e v_H
\end{pmatrix}
 \begin{pmatrix}
 s_{e_R} & c_{e_R} \\ c_{e_R}& -s_{e_R}
 \end{pmatrix}
 = \mathrm{diag}\left(\mu_{E_1}, \mu_{E_2}\right), 
\end{align}
where $\mu_{E_{1,2}}$ are, in general, not positive. 
Under the assumption, $m_L, m_E > 0$ and $v_H \ll m_E$, 
$\mu_{E_a} > 0$, and thus $\mu_{E_a} = m_{E_a}$ given by 
Eq.~\eqref{eq-mEsingApp}. 
} 
\begin{align}
\label{eq-mEsingApp}
 m_{e_1} \simeq&\ 
 y_1 v_H + 
 v_\Phi \eta_e , 
\quad 
 m_{E_1} \simeq %&\ 
\sqrt{\frac{S_e - \sqrt{D_e}}{2}}, 
\quad 
 m_{E_2} \simeq %&\ 
\sqrt{\frac{S_e + \sqrt{D_e}}{2}}, 
\end{align}
where 
\begin{align}
 \eta_e := \la_L \la_E v_\Phi 
 \left( \frac{s_{e_L}c_{e_R}}{m_{E_1}}+ \frac{c_{e_L}s_{e_R}}{m_{E_2}}\right)
\sim \la_e\la_L\la_E \frac{ v_\Phi v_H}{m_L m_E} + \order{\frac{v_H^3}{m_E^3}}. 
\end{align}

For the neutrinos, 
the diagonalization matrices are given by 
\begin{align}
 U_{n_L} = 
\begin{pmatrix}
 1 & 0 & 0 \\ 
 0 & c_{L_1} & s_{L_1} \\ 
 0 & -s_{L_1}& c_{L_1}
\end{pmatrix}
\begin{pmatrix}
 c_{L_2} & s_{L_2} & 0 \\
 -s_{L_2}& c_{L_2} & 0 \\ 
 0 & 0 & 1 
\end{pmatrix}
\begin{pmatrix}
 1 & 0 & 0 \\ 
 0 & c_{n_L} & s_{n_L} \\ 
 0 & -s_{n_L}& c_{n_L}
\end{pmatrix}, 
\quad 
 U_{n_R} = 
\begin{pmatrix}
 s_{n_R} & c_{n_R} \\
 c_{n_R}& -s_{n_R} \\ 
\end{pmatrix}, 
\end{align}
where 
\begin{align}
c_{L_1} := \frac{m_N}{\sqrt{m_N^2 + \la_n^2v_H^2}}, 
\quad 
s_{L_1} := \frac{\la_nv_H}{\sqrt{m_N^2 + \la_n^2v_H^2}}, 
\quad 
c_{L_2} := \frac{c_{L_1}m_L-s_{L_1}\la_n^\pr v_H}{\tilde{m}_L}, 
\quad 
s_{L_2} := \frac{\la_L v_\Phi}{\tilde{m}_L}, 
\end{align}
with $\tilde{m}_L := \sqrt{\la_L^2 v_\Phi^2 + (c_{L_1}m_L-s_{L_1} \la_n^\pr v_H)^2}$. 
The first matrix is to rotate away the $(2,1)$ element, 
and then the $(1,2)$ element is rotated away by the second matrix. 
The angles in the last matrix, $c_{n_{L,R}}$ and $s_{n_{L,R}}$, are given by 
formally replacing $\la_e \to 0$, $m_E \to \sqrt{m_N^2 + \la_n^2 v_H^2}$, 
$m_L \to \tilde{m}_L$ and $\la_e^\pr v_H \to s_{L_1} m_L + c_{L_1} \la_n^\pr v_H$ 
from $c_{e_{L,R}}$ and $s_{e_{L,R}}$ shown in Eq.~\eqref{eq-cseLR}. 
The singular values $m_{N_1}$ and $m_{N_2}$ are respectively obtained 
by the same replacement from $m_{E_1}$ and $m_{E_2}$ in Eq.~\eqref{eq-mEsingApp}. 
Note that the diagonalization for $\Mcal_n$ 
is exact, not relying on any approximation.

For $v_H \ll m_E$ and $m_L < m_E$, the mixing angles are approximately given by 
\begin{align}
s_{e_L} \sim v_H \frac{\la_e m_E+\la_e^\pr m_L}{\abs{m_E^2-m_L^2}}, 
\quad 
s_{e_R} \sim -v_H \frac{\la_e m_L+\la_e^\pr m_E}{\abs{m_E^2-m_L^2}}. 
\end{align}
The masses of the vector-like leptons are given by 
\begin{align}
 m_{E_1} \sim m_L 
 - v_H^2 \frac{(\la_e^2+\la_e^{\pr2})m_L+2\la_e\la_e^\pr m_E}{2(m_E^2-m_L^2)}, 
\quad 
 m_{E_2} \sim m_E 
 + v_H^2 \frac{(\la_e^2+\la_e^{\pr 2})m_E + \la_e\la_e^\pr m_L}{2(m_E^2-m_L^2)}, 
\end{align}
for $m_L < m_E$ and $m_E-m_L \gg v_H$. 
For the neutrinos, 
$\la_n v_H \lesssim m_N \lesssim \order{1}~\GeV$ is necessary 
to make the vector-like neutrino $N_1$ light so that the dark photon can decay. 
We shall assume $\la_n = 0$ for simplicity. 
The neutrino mixing angles are approximately given by 
\begin{align}
 c_{n_L} \sim \frac{\la_n^\pr v_H}{\sqrt{m_L^2 + \la_n^{\pr 2} v_H^2}}, 
\quad 
 s_{n_L} \sim \frac{m_L}{\sqrt{m_L^2 + \la_n^{\pr 2} v_H^2}}, 
\quad 
c_{n_R} \sim 0, 
\quad 
 s_{n_R} \sim -1, 
\end{align}
and the vector-like neutrino masses are given by 
\begin{align}
 m_{N_1} \simeq \frac{m_N m_L}{m_L^2 + \la_n^{\pr 2} v_H^2}, 
\quad 
 m_{N_2} \simeq \sqrt{m_L^2 + \la_n^{\pr 2} v_H^2}. 
\end{align}

\subsection{Lepton couplings}

The approximate forms of $\Ecal_{A}$, $\Ncal_A$ and $h_A$ are given by 
\begin{align}
 \Ecal_L \sim&\ 
\begin{pmatrix}
1- \eta_e^2/\la_E^2 & s_{e_L} \eta_e/\la_E & - c_{e_L} \eta_e/\la_E \\ 
 s_{e_L} \eta_e/\la_E &
 c_{e_L}^2 &
 c_{e_L}s_{e_L} \\ 
 -c_{e_L} \eta_e/\la_E &
 c_{e_L}s_{e_L}& 
 s_{e_L}^2 
\end{pmatrix}, 
%\\ \notag 
\quad 
 \Ecal_R \sim %&\ 
\begin{pmatrix}
\eta_e^2/\la_L^2 & c_{e_R} \eta_e/\la_L & -s_{e_R} \eta_e/\la_L \\ 
c_{e_R} \eta_e/\la_L & c_{e_R}^2 & -c_{e_R}s_{e_R} \\ 
-s_{e_R}\eta_e/\la_L & -s_{e_R}c_{e_R} & s_{e_R}^2 
\end{pmatrix}, 
\\ \notag 
\Ncal_L \sim &\ 
\begin{pmatrix}
1 & 0 & 0 \\ 0 & c_{n_L}^2 & c_{n_L} s_{n_L} \\ 0 & c_{n_L} s_{n_L}& s_{n_L}^2 
\end{pmatrix}, 
\quad 
\Ncal_R = %&\ 
\begin{pmatrix}
 c_{n_R}^2 & -c_{n_R}s_{n_R} \\ -c_{n_R}s_{n_R} & s_{n_R}^2 
\end{pmatrix}, 
\end{align}
\begin{align}
h_L \sim &\ 
\begin{pmatrix}
 1-\frac{1}{2}(\eta_{L_1}^2+\eta_{L_2}^2 + s_{L_2}^2) 
+s_{L_2}(c_{e_L}\eta_{L_1}-s_{e_L}\eta_{L_2}) 
& \eta_{L_1}- c_{e_L}s_{L_2} & -\eta_{L_2} -s_{e_L}s_{L_2} \\
c_{n_L}(s_{L_2}-c_{e_L}\eta_{L_1}+s_{e_L}\eta_{L_2}) & c_{e_L}c_{n_L} &s_{e_L}c_{n_L} \\
s_{n_L}(s_{L_2}-c_{e_L} \eta_{L_1}+ s_{e_L}\eta_{L_2})&c_{e_L}s_{n_L} & s_{e_L}s_{n_L} \\
\end{pmatrix}
, 
\\ \notag
h_R \sim&\ 
\begin{pmatrix}
 c_{n_R}(c_{e_R}\eta_{R_1}+s_{e_R}\eta_{R_2}) &
 c_{e_R}c_{n_R}&
-c_{n_R}s_{e_R} 
\\ 
 - s_{n_R}(c_{e_R}\eta_{R_1}+s_{e_R}\eta_{R_2}) &
 -c_{e_R} s_{n_R} &
s_{e_R}s_{n_R} 
\\ 
\end{pmatrix}, 
\end{align}
up to $\order{\eta^2}$ and $\order{s_{L_2}^2}$. Here, we take $s_{L_1} = 0$ and the sub-dominant contributions in the lower-right $2\times 2$ block
are omitted. 
For the $Z^\prime$-boson couplings, 
\begin{align}
Q^\prime_{e_L}
\sim&\ 
\begin{pmatrix}
\eta_{L_1}^2+\eta_{L_2}^2 & -\eta_{L_1} & \eta_{L_2} \\ 
 -\eta_{L_1} & 1 & 0 \\
 \eta_{L_2} & 0 & 1 \\ 
\end{pmatrix}, 
\quad 
Q^\prime_{e_R}
\sim
\begin{pmatrix}
\eta_{R_1}^2+\eta_{R_2}^2 & \eta_{R_1} & -\eta_{R_2} \\ 
 \eta_{R_1} & 1 & 0 \\
 -\eta_{R_2} & 0 & 1 \\ 
\end{pmatrix}, 
\\ \notag 
Q^\prime_{n_L}
\sim&\ 
\begin{pmatrix}
 s_{L_2}^2 & -c_{n_L} s_{L_2} & -s_{n_L} s_{L_2} \\ 
-c_{n_L} s_{L_2} & 1 & 0 \\ 
-s_{n_L} s_{L_2} & 0 & 1 
\end{pmatrix}, 
\quad 
Q^\prime_{n_R}
=
\begin{pmatrix}
 1 & 0 \\ 0 & 1 
\end{pmatrix}. 
\end{align}
The Yukawa couplings are given by 
\begin{align}
 Y_e^\chi 
\sim&\ 
\begin{pmatrix}
2\eta_e & 
 c_{e_R}\la_L & -s_{e_R} \la_L \\ 
 -s_{e_L} \la_E & 
 \la_E s_{e_L}\eta_{R_1} + \la_L c_{e_R}\eta_{L_1} & 
-\la_E s_{e_L}\eta_{R_2} - \la_L s_{e_R}\eta_{L_1} \\ 
c_{e_L}\la_E & 
-\la_E c_{e_L} \eta_{R_1} - \la_L c_{e_R}\eta_{L_2} & 
 \la_E c_{e_L} \eta_{R_2} + \la_L s_{e_R}\eta_{L_2} \\ 
\end{pmatrix}, 
\\ \notag 
 Y_n^\chi \sim&\ \la_L 
\begin{pmatrix}
 c_{n_R} & -s_{n_R} \\ 
 c_{n_L} c_{n_R} s_{L_2} & - c_{n_L} s_{n_R} s_{L_2} \\ 
 s_{n_L} c_{n_R} s_{L_2} & - s_{n_L} s_{n_R} s_{L_2} \\
\end{pmatrix}. 
\end{align}

\section{Loop functions} 

\label{app-loopfun}
 
The loop functions for $\Delta a_\ell$ are given by 
\begin{align}
 F_Z(x) =\frac{5x^4-14x^3+39x^2-38 x+8-18x^2 \ln{(x)}}{12(1-x)^4}, \quad %\\
 G_Z(x) =\frac{x^3+3x-4-6x \ln{(x)}}{2(1-x)^3},
\end{align}
and 
\begin{align}
 F_S(y) = -\frac{y^3-6y^2+3y+6y \ln{(y)} + 2}{6(1-y)^4}, \quad
 G_S(y) = \frac{y^2-4y+2\ln{(y)}+3}{(1-y)^3}.
\label{eq-GS}
\end{align}
The loop functions for the oblique parameters are given by 
\begin{align}
 \theta_+(y_1,y_2) = y_1 + y_2 - \frac{2y_1y_2}{y_1-y_2} \log\frac{y_1}{y_2}, 
\quad 
 \theta_-(y_1,y_2) = 2\sqrt{y_1y_2} \left( \frac{y_1+y_2}{y_1-y_2} \log \frac{y_1}{y_2}-2 \right), 
\end{align}
\begin{align}
 \chi_+(y_1,y_2) =&\ \frac{y_1+y_2}{2}-\frac{(y_1-y_2)^2}{3} + 
 \left(\frac{(y_1-y_2)^3}{6}-\frac{1}{2}\frac{y_1^2+y_2^2}{y_1-y_2}
 \right) \log \frac{y_1}{y_2} \\ \notag 
 &\ + \frac{y_1-1}{6} f(y_1,y_1) + \frac{y_2-1}{6} f(y_2,y_2) 
 + \left(\frac{1}{3}-\frac{y_1+y_2}{6}-\frac{(y_1-y_2)^2}{6} \right) f(y_1,y_2), \\
\chi_-(y_1,y_2) =&\ 
 -\sqrt{y_1y_2} \left[ 
 2+\left(y_1-y_2-\frac{y_1+y_2}{y_1-y_2} \right)\log\frac{y_1}{y_2} 
 + \frac{f(y_1,y_1)+f(y_2,y_2)}{2} - f(y_1,y_2) 
\right], 
\end{align}
and 
\begin{align}
 \psi_+(y_1, y_2) =&\ 
 \frac{2y_1+10y_2}{3} + \frac{1}{3} \log\frac{y_1}{y_2}+ 
 \frac{y_1-1}{6} f(y_1,y_1) + \frac{5y_2 + 1}{6} f(y_2,y_2), \\ 
 \psi_-(y_1,y_2) =&\ 
 - \sqrt{y_1 y_2}\left(4+\frac{f(y_1,y_1)+f(y_2,y_2)}{2} \right). 
\end{align}
Here, the function $f$ is defined as 
\begin{align}
\label{eq-deff}
 f(y_1,y_2) = 
\begin{cases}
 \sqrt{d} \log\abs{\dfrac{y_1+y_2-1+\sqrt{d}}{y_1+y_2-1-\sqrt{d}}} & d > 0 \\ 
 0 & d=0 \\ 
 -2 \sqrt{\abs{d}} \left[\tan^{-1} \dfrac{y_1-y_2+1}{\sqrt{\abs{d}}}-
 \tan^{-1} \dfrac{y_1-y_2-1}{\sqrt{\abs{d}}}\right] & d < 0 
\end{cases}
\end{align}
with $d = (1+y_1-y_2)^2 -4y_1$.

%%%%%%%%%%%%%%%%%%%%%%%%%%%%%%%%%%%%%%%%%%%%%%%%
{\small
\bibliographystyle{JHEP}
\bibliography{ref}}
%%%%%%%%%%%%%%%%%%%%%%%%%%%%%%%%%%%%%%%%%%%%%%%%
\end{document}